\documentclass[lettersize,journal]{IEEEtran}

\usepackage{cite}				
\usepackage{graphicx}	
\usepackage{amsmath}		
\usepackage{algorithmic}	
\usepackage{array}		%
\usepackage{dsfont}	
\ifCLASSOPTIONcompsoc	
\usepackage[caption=false,font=normalsize,labelfont=sf,textfont=sf]{subfig}
\else
\usepackage[caption=false,font=footnotesize]{subfig}
\fi
\usepackage{stfloats}
\usepackage{url}
\usepackage{algorithm}	
\usepackage{color}		
\usepackage[dvipsnames]{xcolor}
\usepackage{bm}			
\usepackage{amsfonts}	
\usepackage{amssymb}	
\usepackage{booktabs}   
\usepackage{makecell}   
\usepackage[stable]{footmisc}
\usepackage{multirow}
\usepackage[switch]{lineno}	
\usepackage{dsfont}	    
\usepackage{textcomp}

\newtheorem{problem}{Problem}

\begin{document}

\title{Successful Transmission Probability and SIR Meta Distribution Analysis for Multi-Antenna Cache-Enabled Networks with Interference Nulling}

\author{Tianming~Feng, Chenyu~Wu, Xiaodong~Zheng, Peilin~Chen, Yilong~Liu, and Shuai~Han,~\IEEEmembership{Senior~Member,~IEEE}	
	\thanks{T. Feng, X. Zheng, P. Chen, and Y. Liu are with the 54th Research Institute of China Electronics Technology Group Cooperation, Shijiazhuang 050081, China (e-mails: feng$\_$tm@163.com, zheng$\_$xiaodong@sohu.com, chen$\_$etc@163.com, lylogn@126.com).}
	\thanks{C. Wu and S. Han are with the School of Electronic and Information Engineering, Harbin Institute of Technology, Harbin 150001, China (e-mail: \{wuchenyu, hanshuai\}@hit.edu.cn).}
\vspace{-2em}}	


\maketitle

\begin{abstract}
This paper investigates a multi-antenna cache-enabled network with interference nulling (IN) employed at base stations. Two IN schemes, namely, the fixed IN scheme and the flexible IN scheme are considered to improve the received signal-to-interference ratio (SIR) at users. To thoroughly explore the effects of the caching parameter and the IN parameters on the network performance, we focus on the analysis of not only the successful transmission probability (STP) but the SIR meta distribution. For each IN scheme, the expression for the STP is derived and an approximated expression for the SIR meta distribution is also obtained by deriving the first and second moments of an upper bound of the link reliability and utilizing the beta distribution. With this analytical framework, we compare the performance of these two IN schemes and gain some useful system design guidelines from the perspectives of the STP and the SIR meta distribution by numerical simulations.
\end{abstract}


\section{Introduction}\label{sec1}

\subsection{Motivation}
\IEEEPARstart{C}{ache-enabled} networks (CENs), where base stations (BSs) are equipped with cache to pre-fetch popular files during the peak-off time, have been proved to be an effective solution to alleviate the backhaul burden caused by the exponential increase of the mobile data traffic \cite{Dimakis2013}, as well as reduce the transmission delay \cite{Li2017a}. However, the strong interference suffered by users is a noteworthy problem in CENs, since the serving BS of a user may not be its geographically nearest BS due to the limited cache size at BSs \cite{Liu2016a}. To improve the quality of received signals at users, multiple antennas can be deployed at each BS in CENs to either boost the user-received desired signals by providing extra spatial diversity \cite{Kuang2019,Kuang2019a}; or mitigate the interference at each user by conducting interference coordination \cite{Jiang2019c,Xu2019c,Zhi2019,Liu2021,Feng2022}. 

However, the performance metrics considered in the aforementioned works are all based on the successful transmission probability (STP), which is the complementary cumulative distribution function (CCDF) of the signal-to-interference ratio (SIR) at the typical user and is obtained by taking an expectation over the point process. Even though the STP is of importance for reflecting the average performance of the networks, it does not provide any information on the distribution of the performance of individual links for each network realization.

To this end, an innovative performance metric called the \textit{meta distribution} of the SIR is proposed in \cite{Haenggi2016}, which can provide more fine-grained information by characterizing the distribution of the conditional STP (CSTP) conditioning on the realization of BS point process. The meta distribution can answer the question ``given the randomness of the locations of BSs, what is the distribution of the probability of the event that the SIR threshold is met for any user?'', which reflects the achievable performance for a given percentile of users,\footnote{For example, if we focus on the 5th-percentile, the meta distribution gives the link reliability that $95\%$ of the users can achieve.} and is an important design criterion for the network operators; whereas the STP only answers the question ``with a given SIR threshold, what fraction of users can achieve successful transmission on average?" \cite{Haenggi2016}.

So far, there has not been any work focusing on the meta distribution analysis on multi-antenna CENs, not to mention the interference nulling (IN). The works in \cite{Jiang2019c,Xu2019c,Zhi2019,Liu2021,Feng2022} with IN only study the average performance of the networks. However, from the perspective of system design, it is important to evaluate the network performance from the aspects of not only the average STP, but also the link reliability distribution. When considering IN in multi-antenna CENs, the caching parameter and the IN parameters are coupled with each other in a complicated manner, the effects of these two groups of parameters on the distribution of link reliability are still not clear and need further investigation. This motivates the work of this paper. 

\subsection{Contributions}
In this work, we investigate multi-antenna CENs with two IN schemes, namely the \textit{fixed IN scheme} and the \textit{flexible IN scheme}, considered. We endeavor to analyze the performance of these two schemes from the perspectives of not only the STP but the SIR meta distribution and further explore the effects of the caching parameters and the IN parameters on these two metrics, so as to obtain more system design insights. Specifically, for both schemes, multiple antennas are equipped at each BS, a part of the spatial degrees-of-freedom (DoF) is allocated for IN, and the remaining DoF is reserved to provide spatial diversity gain. The IN ranges in the fixed IN scheme are the same for all users; whereas the IN ranges for different users in the flexible IN scheme are flexible and depend on their average received signal power. The contributions of this work are summarized as follows.

$\bullet$ We first present a tractable analytical framework for the aforementioned multi-antenna CENs with a flexible uniform distribution caching policy considered under the fixed IN scheme and the flexible IN scheme.  Using stochastic geometry analysis techniques, we obtain tractable expressions for the STPs under both IN schemes. These expressions enable us to efficiently analyze the impacts of the caching parameter (i.e., the file diversity gain) and the IN parameters (i.e., the IN range and the maximum IN DoF) on the average network performance, i.e., the STP.

$\bullet$ We derive a tight upper bound on the CSTP, with which we further provide approximated expressions for the SIR meta distributions of the two IN schemes. Specifically, due to the difficulty of attaining the exact expression for the distribution of the CSTP, we focus on the upper bound, derive the first and second moments of this upper bound, and utilize the beta distribution to obtain approximated expressions for the SIR meta distributions, which facilitates the investigation into the effects of the caching parameter and the IN parameters on the distribution of link reliability.

$\bullet$ With the aforementioned analytical framework and via numerical simulations, for the low and high SIR threshold regions, we provide several system design guidelines for the purposes of improving the STP, improving the link fairness, and increasing the fraction of users with high link reliability. Moreover, we further provide a numerical optimization framework to explore more design guidelines quantitatively.

The rest of this paper is organized as follows. A literature survey is presented in Section \ref{section: Meta Related Works}. The system model is illustrated in Section \ref{section: Meta System}. Section \ref{section: Meta Auxilliary Results} provides the analytical results for the STP and SIR meta distribution. The numerical results are provided in Section \ref{section: Meta Numerical}, and the conclusions are drawn in Section \ref{section: Meta Conclusion}.

\section{Related Works}\label{section: Meta Related Works}
\subsection{Interference Management for Cache-Enabled Networks}
In cache-enabled networks, users can suffer severe interference due to the content-centric user association mechanism, when their serving BSs are not the geographically nearest BSs. To tackle this problem, some interference management techniques are considered in \cite{Chae2017,Wen2018a,Feng2021,Kuang2019,Kuang2019a,Jiang2019c,Xu2019c,Zhi2019,Liu2021,Feng2022}. Specifically, in \cite{Chae2017,Wen2018a,Feng2021}, several different BS cooperation transmission schemes for different network scenarios are proposed to enhance the received signal power at each user. However, they only focus on the single-antenna layout in the network, the advantages of multi-antenna are not fully exploited.

Multi-antenna technique is also an effective method to improve the received signal quality at users in CENs \cite{Kuang2019,Kuang2019a,Jiang2019c,Xu2019c,Zhi2019,Liu2021,Feng2022}. Specifically, by deploying multiple antennas at each BS, \cite{Kuang2019} and \cite{Kuang2019a} take advantage of the spatial diversity gain it brings to strengthen the desired signal at each user in multi-tier and limited-backhaul network scenarios, respectively; whereas \cite{Jiang2019c,Xu2019c,Zhi2019,Liu2021,Feng2022} utilize the available spatial DoF for IN so as to suppress the interference suffered by users. In \cite{Jiang2019c}, multiple antennas are equipped at each user, and the IN is implemented on the receiver side. The authors in \cite{Xu2019c} divide the multi-antenna BSs into clusters and use coordinated beamforming to cancel the intra-cluster interference of each BS at the users within the cluster. Starting from this cluster-based IN scheme, the authors in \cite{Liu2021} further consider multiple antennas equipped at each user to form a multiple-input multiple-output (MIMO) system, so that more users can simultaneously enjoy an interference-free link in each cluster with interference alignment applied. In addition, the works \cite{Zhi2019} and \cite{Feng2022} investigate more flexible user-centric IN schemes with a fixed IN range and a flexible IN range, respectively; and the per-user throughput and the area spectrum efficiency are investigated. Note that all the works above only focus on the metrics about the averaged performance. Therefore, no system design insight from the perspective of individual links can be obtained.

\subsection{Successful Transmission Probability Based Analysis}
The STP, also known as the coverage probability in some works, is an important metric to evaluate the average link reliability in wireless networks. For the works \cite{Jiang2019c,Xu2019c,Liu2021,Feng2022} considering IN in multi-antenna CENs, the STP serves as either a main performance metric to be maximized \cite{Jiang2019c}, or an intermediate metric for investigating other metrics of interest, such as the hit probability \cite{Xu2019c,Liu2021} and the spectral efficiency \cite{Xu2019c,Feng2022}. To tackle the high-order derivatives of the Laplace transform of the interference when analyzing the STP in multi-antenna networks, the authors in \cite{Jiang2019c} resort to the Faa di Bruno's formula, which leads to a result with complicated form; the authors of \cite{Xu2019c} and \cite{Liu2021} consider approximated expressions for STP to avoid the calculation of high-order derivatives; whereas our previous work \cite{Feng2022} derives an exact expression for the STP using the $L_1$-induced norm of a Toeplitz matrix with simple structure. In this paper, we follow the work in \cite{Feng2022} to conduct further investigation of the STP. By introducing a parameter called the maximum IN DoF, we have more freedom to analyze the impact of IN and spatial diversity on the STP as well as the distribution of individual link performance, which, however, is not considered in \cite{Feng2022}.

\subsection{Meta Distribution Based Analysis}
As mentioned before, the meta distribution can provide more detailed information on per-link performance, the concept of which is first introduced in \cite{Haenggi2016}. Following the work \cite{Haenggi2016}, the meta distribution is investigated in various network scenarios, such as the Poisson cellular networks with BS cooperation \cite{Cui2018a,Feng2019}; heterogeneous networks (HetNets) with joint spectrum allocation and user offloading \cite{Deng2019}, and cell range expansion \cite{Wang2019g}; non-Poisson HetNets \cite{Kalamkar2019}; device-to-device (D2D) communication underlaying cellular networks \cite{Salehi2017}, and millimeter-wave D2D networks \cite{Deng2017}; non-orthogonal multiple access (NOMA) networks \cite{Salehi2019} for both uplink and downlink; and cache-enabled networks with random discontinuous transmission \cite{Yang2022}. However, none of the aforementioned works studies the multi-antenna scenario. 

While the work \cite{Wang2018c} provides SIR meta distribution analysis with interference cancellation taken into consideration, it only considers an ideal scheme called the close interference cancellation, where the effectiveness of the scheme is simply reflected by a coefficient, and the wireless channel characteristic brought by multiple antennas is not considered. When implementing multi-antenna techniques for IN in cache-enabled networks, it remains unknown the effects of the IN range, the DoF allocated to IN, and the file diversity gain on the distribution of the individual links and the fairness among different users.

In this paper, we will shed some light on the above issues. We consider two IN schemes, i.e., the fixed IN scheme and the flexible IN scheme, and compare the network performance of these two schemes from the perspectives of not only the STP but the SIR meta distribution, so as to provide more fine-grained information and obtain more system design insights.

\section{System Model}\label{section: Meta System}
\subsection{Network and Caching Model}
We consider a single-tier cache-enabled multi-antenna network. The locations of base stations (BSs) are modeled as a homogeneous Poisson point process (PPP) in $\mathbb{R}^2$, denoted by $\Phi$ with density $\lambda$. Each BS is equipped with $M$ antennas and is considered to serve one user over one time-frequency resource block with transmit power $P$. Single-antenna users are also distributed as a homogeneous PPP $\Phi_u$ with density $\lambda_u$. We consider the downlink transmission and assume a full-loaded scenario so that each BS always has at least one user connected to it. Without loss of generality, we assume a typical user $u_0$ located at the origin and focus on the performance analysis of this user according to Slivnyak's theorem \cite{Haenggi2012}. 

Let $\mathcal{N} = \{1,2,\cdots,N\}$ be a content library containing $N$ different files. All files are assumed to have the same size which equals 1. For any given user, the probability that file $n$ is requested is $a_n \in [0,1]$, which is called the file popularity, and we have $\sum_{n \in \mathcal{N}} a_n = 1$. Thus, the file popularity distribution is given by $\mathbf{a} \triangleq (a_n)_{n\in \mathcal{N}}$. Note that the popularity distribution changes slowly over time, and can be estimated using learning-based methods \cite{Garg2019,Jiang2019d}. Therefore, we assume it is known as a priori and remains unchanged over the period of interest. Without loss of generality, we assume that a file of smaller index has higher popularity, i.e., $a_1 > a_2 > \cdots > a_N $. Each BS is equipped with a cache of size $C \leq N$ and can store $C$ different files.

In this paper, we consider a \textit{flexible uniform distribution caching} (FUDC) policy. Specifically, consider a tunable parameter $\xi^\prime \in [1, \frac{N}{C} ]$. With this parameter, the $N_c =\lfloor \xi^\prime  C \rfloor $ most popular files from $\mathcal{N}$ are chosen to form a file set $\mathcal{N}_c = \{1,2,\cdots,N_c \}$, where $\lfloor x \rfloor$ denotes the rounding down of $x$. All the files in $\mathcal{N}_c$ can be stored in the network, and each BS will uniformly randomly select $C$ different files out of $\mathcal{N}_c$ to store, with the same caching probability $T_c = \frac{C}{ N_c }$. We refer to the discrete parameter $\xi = \frac{N_c}{ C}$ as the \textit{file diversity gain}. Therefore, FUDC is a combination of the most popular caching (MPC) policy \cite{Bastug2015} and the uniform distribution caching (UDC) policy \cite{Tamoor-ul-Hassan2015a}. When setting $\xi = 1$, the $N_c=C$ most popular files from $\mathcal{N}$ are cached at each BS with probability $T_c = 1$, which is exactly the MPC policy; whereas when setting $\xi = \frac{N}{C} $, we have $\mathcal{N}_c = \mathcal{N}$, and each BS will randomly select $C$ files out of $\mathcal{N}_c$ to store, according to the uniform distribution, which is the same as the UDC policy. The FUDC policy takes both file popularity and file diversity into consideration with only one parameter $\xi$. In particular, by adjusting the value of $\xi$, a high cache hit probability can be ensured for the $N_c$ most popular files; whereas randomly storing these $N_c$ files at each BS makes the spatial file diversity fully exploited. Even though in some scenarios, the performance of this policy may not be as good as that of the optimal policies studied in \cite{Jiang2019c,Kuang2019,Xu2019c}, the simplicity of FUDC will significantly facilitate the analysis of the network performance.

For the user association, consider the content-centric association mechanism \cite{Cui2016}, where a user will connect to its nearest BS storing the file it requests. This BS is referred to as its serving BS and can provide the maximum long-term average receive power for the requested file. In this case, the serving BS of a user may not be its geographically nearest BS, thus the interference caused by the non-serving BSs that are closer to the user than its serving BS will significantly deteriorate the quality of the received signal at the user. This motivates us to use user-centric IN schemes to suppress the interference received by the user, which will be elaborated on later. When the requested files of users are not stored in the network, BSs may retrieve these uncached files from the core network via backhaul links. The investigation of this case is beyond the scope of this paper, thus we omit it as in \cite{Cui2016,Wen2017}, and simply regard it as a failed transmission.

\subsection{Interference Nulling Schemes}\label{subsec: Meta IN scheme}
In this paper, we consider two interference nulling schemes for analysis, namely, the \textit{fixed IN scheme} and the \textit{flexible IN scheme}. In both schemes, each BS equipped with $M$ antennas randomly selects one user out of its all potential users to serve. To mitigate the strength of received interference, each served user will send an IN request to all the interfering BSs within a given distance. This area is referred to as the \textit{IN range}. Consider a parameter $L\in \mathbb{N}$ called the \textit{maximum IN DoF}. Each BS reserves $L \in [0, M-1]$ antennas to suppress its interference at up to $L$ other users and the remaining DoF is used to provide spatial diversity to boost the desired signal power at its served user.

The main difference between these two IN schemes lies in the selection of the IN range for each user. Specifically, for the fixed IN scheme, the IN range for each user is set to be a given value, denoted by $R_c$. Whereas, for the flexible IN scheme, a tunable parameter $\mu \in [0,+\infty)$, which is referred to as the \textit{IN coefficient}, is chosen to control the IN range of each user. Suppose the distance between a user and its serving BS is $Z$, which is referred to as the \textit{serving distance}. The user will send an IN request to all the interfering BSs within the circle of radius $\mu Z$, which is the IN range of the flexible IN scheme. Note that for the flexible IN scheme, the IN ranges for different users are generally different, since the serving distances are usually different. Therefore, the flexible IN scheme takes the different interference situations for different users into consideration, which, however, is not considered in the fixed IN scheme, where the IN ranges are set to be equal for all users.	

For both IN schemes, when a BS receives IN requests, it will suppress its interference at up to $L$ requesting users, due to the limitation of available antennas. Let $\varTheta_r $ denote the number of IN requests received by a BS. If $\varTheta_r $ is greater than $L$, the BS uniformly randomly selects $L$ requests to satisfy; otherwise, the BS cancels its interference at all $\varTheta_r$ users, and the extra DoF is used to provide its served user with a spatial diversity order $M- \varTheta_r $. 

For the fixed IN scheme, we need to consider the relation between the serving distance $Z$ and the IN range $R_c$. When $ R_c < Z$ (resp. $R_c \geq Z$), all the interfering BSs of $u_0$ consist of: 1) the interfering BSs within the circle of radius $R_c$ (resp. $Z $); 2) the interfering BSs within the annulus from radius $R_c $ (resp. $Z $) to $Z $ (resp. $ R_c $); and 3) the remaining interfering BSs outside the circle of radius $Z $ (resp. $ R_c $). Note that the interference within the circle of radius $Z$ comes from those BSs that do not store the file requested by $u_0$ and whose interference is not nulled out. Denote by $\Phi_a^{\text{fx}}$, $\Phi_b^{\text{fx}}$, and $\Phi_c^{\text{fx}}$ the sets of interfering BSs corresponding to the above three cases. We have 
\begin{equation}\label{equ: Meta Phi a b c Rc}
\Phi_{i}^{\text{fx}} \triangleq \{ x | x \in \Phi \backslash \{ x_{0}\} , \| x \|\in \varOmega_{i}^{\text{fx}} \}, \quad i\in\{a,b,c\},
\end{equation}
where $x_0$ denotes the serving BS of $u_0$, and
\begin{align}\label{equ: Meta Omega abc Rc}
\varOmega_a^{\text{fx}} &= [0,  \min \{Z, R_c\}), \notag
\\ \varOmega_b^{\text{fx}} &= [ \min \{ Z, R_c \}, \max \{Z, R_c \}), \notag
\\ \varOmega_c^{\text{fx}} &= [ \max \{Z, R_c \}, +\infty),
\end{align}
are three distance intervals.

For the flexible IN scheme, we need to consider the cases that $0 \leq \mu <1$ and $\mu \geq 1$. Similarly to the fixed IN scheme, the interfering BSs can also be categorized into three groups, denoted by $\Phi_a^{\text{fl}}$, $\Phi_b^{\text{fl}}$, and $\Phi_c^{\text{fl}}$ with 
\begin{equation}\label{equ: Meta Phi a b c}
\Phi_{i}^{\text{fl}} \triangleq \{ x | x \in \Phi \backslash \{ x_{0}\} , \| x \|\in \varOmega_{i}^{\text{fl}} \}, \quad i\in\{a,b,c\},
\end{equation}
and
\begin{align}\label{equ: Meta Omega abc}
\varOmega_a^{\text{fl}} &= [0,  \min \{Z, \mu Z \}), \notag
\\ \varOmega_b^{\text{fl}} &= [ \min \{ Z, \mu Z \}, \max \{Z, \mu Z \}), \notag
\\ \varOmega_c^{\text{fl}} &= [ \max \{Z, \mu Z \}, +\infty).
\end{align}

\subsection{SIR Model}\label{subsec: Meta SIR Model}
In this paper, we assume perfect channel state information (CSI) and implement the low-complexity zero-forcing beamforming (ZFBF) at each BS to serve its user as well as suppress its interference at other users. Note that ZFBF is a commonly used technique for interference management, and provides a reasonable balance between the performance and tractability \cite{Dhillon2013,Li2015a,Hosseini2018}. We leave more other precoding techniques, e.g., minimum mean square error (MMSE), for future work.

For the wireless channel, we consider the standard power-law path loss model and Rayleigh fading coefficient. Therefore, the received power at the typical user $u_0$ from a BS located at $x \in \Phi $ is given by $P g_x \|x \|^{-\alpha}$, where $\alpha >2$ is the path loss exponent and $g_x$ is the channel gain. With ZFBF and perfect CSI, when we assume the Rayleigh fading channel, if $x$ is the serving BS, $g_x$ denotes the effective channel gain of the desired signal, and follows Gamma distribution, i.e., $g_{x} \overset{d}{\sim} \Gamma(M - \min \{ \varTheta_{r}, L\}, 1)$ \cite{Dhillon2013}; however, if $x$ is an interfering BS, $g_x$ denotes the interfering channel gain between $u_0 $ and BS $x$, and follows $g_{x} \overset{d}{\sim} \Gamma(1, 1) = \text{Exp}(1)$, which is the exponential distribution \cite{Dhillon2013}. 

Denote by $x_0$ the serving BS of $u_0$. Then, when the file $n$ is requested, the received signal to interference ratio (SIR) at $u_0$ is given by
\begin{equation}\label{equ: Meta SIR expression}
\Upsilon_{n}=\frac{  g_{x_0 } Z ^{ -\alpha }}{ \sum_{x \in \Phi_a \cup \Phi_b\cup \Phi_c }    g_{x} \|x\|^{-\alpha } },
\end{equation}
where $\Upsilon_{n} $ can be $\Upsilon_{n}^\text{fx}$ or $\Upsilon_{n}^\text{fl}$ corresponding to the fixed IN scheme or flexible IN scheme; $Z$ is the serving distance of $u_0$; $\Phi_i $, $i\in\{a,b,c\} $ can be $\Phi_i^{\text{fx}} $ in \eqref{equ: Meta Phi a b c Rc} or $\Phi_i^{\text{fl}} $ in \eqref{equ: Meta Phi a b c}. Note that in this paper, we investigate SIR, rather than the signal to interference and noise ratio (SINR), since for nowadays densely deployed wireless networks, the power of interference received at users is much larger than that of environment noise so that we can safely neglect the noise.

\subsection{Meta Distribution}
The transmission of a given file $n$ is considered to be successful if the received SIR $\Upsilon_{n}$ at the requesting user is greater than some predefined threshold $\tau$, the probability of which is referred to as the \textit{successful transmission probability} (STP) of this file and is given by
\begin{equation}\label{equ: Meta standard STP n}
p_{s,n} \triangleq  \mathbb{P} [ \Upsilon_{n} \geq \tau ].
\end{equation}	
Furthermore, if the locations of BSs $\Phi$ in a network realization are given, the conditional successful transmission probability (CSTP, or equivalently, the link reliability, i.e., the probability that the wireless channel under consideration gives an SIR exceeding $\tau$ \cite{Haenggi2016}) of a link for transmitting file $n$ is given by
\begin{equation}\label{equ: Meta CSTP n}
P_{s,n} (\tau) \triangleq \mathbb{P}\left[ \Upsilon_{n} \geq \tau \mid \Phi \right].
\end{equation} 

The SIR meta distribution of file $n$ is defined as \cite{Haenggi2016}
\begin{equation}\label{equ: Meta Definition of Meta}
\bar{F}_{P_{s,n}}(x) \triangleq \mathbb{P} \left[ P_{s,n} (\tau) > x \right], \quad x \in[0,1],
\end{equation}
which is the complementary cumulative distribution function (CCDF) of the CSTP $P_{s,n} (\tau)$; $x$ here is some predefined link reliability threshold. Since the point processes $\Phi$ and $\Phi_u$ are ergodic, $\bar{F}_{P_{s,n}}(x)$ can reflect the fraction of users whose link reliability, i.e., $ P_{s,n} (\tau)$, is greater than $x$ in each network realization when file $n$ is requested. Moreover, we note that the STP in \eqref{equ: Meta standard STP n} is the mean of the CSTP in \eqref{equ: Meta CSTP n}. Since $ P_{s,n} (\tau)$ is non-negative, we have the following relation between the STP and the SIR meta distribution:
\begin{equation}\label{equ: Meta Relation bt STP and Meta}
p_{s,n}= \mathbb{E} \left[ P_{s,n} (\tau) \right] = \int_{0}^{1} \bar{F}_{P_{s,n} }(x) \mathrm{d} x.
\end{equation}

From \eqref{equ: Meta Relation bt STP and Meta}, we observe that compared with the STP, the SIR meta distribution provides more detailed information on link reliability given the network realization. From the meta distribution, we can easily obtain the STP; while with an STP, we cannot observe any information about the distribution of the CSTP. In this paper, we pay more attention to the investigation of the SIR meta distribution and the comparison between the STP and the SIR meta distribution, so as to provide more insights into the impact of system design parameters, i.e., $(R_c, L, \xi)$ (for the fixed IN scheme) or $(\mu, L, \xi)$ (for the flexible IN scheme), on the link reliability in the network.

Considering the total probability theorem, the whole STP and the whole SIR meta distribution for the network, denoted by $p_{s} $ and $\bar{F}_{P_{s}}(x)$, are respectively given by
\begin{align}
p_{s} &= \sum_{ n \in \mathcal{N}_c} a_n p_{s,n} , \label{equ: Meta definition of whole STP}
\\\bar{F}_{P_{s} }(x) & = \sum_{ n \in \mathcal{N}_c} a_n \bar{F}_{P_{s,n}}(x).\label{equ: Meta definition of whole Meta}
\end{align}

\subsection{Notation}
In this paper, we use $(\mathsf{R_I}, L,\xi) $ to uniformly denote the design parameters for the two IN schemes, with $ \mathsf{R_I} = R_c$ for the fixed IN scheme and $ \mathsf{R_I} = \mu$ for the flexible IN scheme. For a given variable $X$, $X^\text{fx}$ (resp. $X^\text{fl}$) denotes the variable for the fixed (resp. flexible) IN scheme. If the superscript ``fx'' or ``fl'' is not specified, then $X$ denotes the variable for both IN schemes. Moreover, $X^u$ denotes the upper bound of $X$, and $X_n$ denotes the corresponding variables when file $n$ is requested. All these superscripts and subscripts can combine with each other to endow a variable with different meanings.

\section{The Successful Transmission Probability and SIR Meta Distribution}\label{section: Meta Auxilliary Results}
In this section, we first present some auxiliary results which are necessary for the STP and SIR meta distribution analysis, including the probability mass function (PMF) of the number of IN requests received by a BS and the IN missing probability (whose definition will be given later). Based on these results, the expression for the STP of each IN scheme is obtained. Then, we give upper bounds on the first and second moments of the CSTP of each IN scheme, based on which, an approximation of the SIR meta distribution is then obtained.

\subsection{Auxiliary Results}
\subsubsection{PMF of the Number of IN Requests Received by a BS}	
To facilitate the derivation of the SIR meta distribution, we first need to calculate the PMF of the number of IN requests received by the serving BS of a user, i.e., $\varTheta_r $. We have the following assumptions to make the performance analysis more feasible: 1) the served users form a homogeneous PPP $\tilde{\Phi}_u$, which is a thinning of $\Phi_u$, and is independent from the locations of BSs $\Phi$; and 2) the numbers of IN requests received by different BSs are independent. Similar assumptions are adopted in \cite{Li2015a,Cui2016a}, and the accuracy will be verified later in Section~\ref{section: Meta Numerical}. Based on theses assumptions, $\varTheta_r $ can be approximated by a Poisson random variable with PMF
\begin{equation}\label{equ: Meta PMF theta}
\mathbb{P} \left[ \varTheta_r  = \theta_r \right] \approx \frac{ \bar{\varTheta}  (\mathsf{R_I}, \xi)  ^{\theta_r }}{ \theta_r !} \exp\left(  -\bar{\varTheta}  (\mathsf{R_I}, \xi) \right),
\end{equation}
where $\bar{\varTheta}(\mathsf{R_I}, \xi)$ represents $\bar{\varTheta}^\text{fx} (R_c, \xi)$ or $\bar{\varTheta}^\text{fl} (\mu, \xi) $ corresponding to $\mathsf{R_I} = R_c$ or $\mathsf{R_I} = \mu$, which is the mean of the number of IN requests received by a BS for the fixed IN scheme or the flexible IN scheme and is approximately given by
\begin{align}\label{equ: Meta meanTheta}
\bar{\varTheta}^\text{fx} (R_c, \xi)& \approx \pi \lambda R_c^2 + e^{-\pi \frac{\lambda} {\xi} R_c^2} - 1 ,\notag
\\
\bar{\varTheta}^\text{fl} (\mu, \xi)&  \approx  \xi \mu^2  - \min \{ \mu^2, 1 \}.		
\end{align}
The detailed derivation can be found in Appendix~\ref{appendix: Meta varThetabar}.

From \eqref{equ: Meta meanTheta}, we observe that for the flexible IN scheme, the mean of the number of IN requests received by a BS is only affected by the IN coefficient $\mu$ and the file diversity gain $\xi$; whereas for the fixed IN scheme, the BS density $\lambda$ also has an impact on it besides the IN range $R_c$ and the file diversity gain $\xi$.

\subsubsection{IN Missing Probability}
When the number of IN requests received at a BS is smaller than the maximum IN DoF $L$, all the requests will be satisfied; otherwise, the BS will randomly choose $L$ users to suppress its interference. Therefore, the PMF of the number of IN requests \textit{satisfied} by a BS, for both IN schemes, denoted by $\varTheta_I \in \{0,1,\cdots,L\}$, is given by
\begin{align}\label{equ: Meta PMF of ThetaI}
\mathbb{P} \left[ \varTheta_I \!= \! \theta \right] \! = \!  \left\{
\begin{array}{ll}
\mathbb{P} \left[ \varTheta_r = \theta \right], &  \, 0 \leq \theta \leq L \!- \! 1, \\
\sum_{\theta^\prime = L}^{\infty} \mathbb{P} \left[ \varTheta_r = \theta^\prime \right], &\, \theta = L.\\
\end{array}
\right.
\end{align}
Note that for the case $\theta = L$, we have the following equality to simplify the calculation, i.e., $ \sum_{\theta^\prime = L}^{\infty} \mathbb{P} \left[ \varTheta_r = \theta^\prime \right] = \frac{\gamma( L , \bar{\varTheta} (\mathsf{R_I}, \xi) ) }{\Gamma (L )} $, where $\gamma (s,x)=\int _{0}^{x}t^{s-1}\,\mathrm {e} ^{-t}\,{\rm {d}}t$ is the lower incomplete Gamma function, and $\bar{\varTheta} (\mathsf{R_I}, \xi)$, with $\mathsf{R_I} = \mu$ or $\mathsf{R_I} = R_c$ is given by \eqref{equ: Meta meanTheta}.

Then, as we show in Appendix~\ref{appendix: Meta epsilon}, the following formula \eqref{equ: Meta epsilon} gives the probability that a BS receives an IN request from the typical user $u_0$ but fails to null the interference at it, which is referred to as the \textit{IN missing probability}, and is denoted by $\varepsilon (\mathsf{R_I}, L,\xi) $.
\begin{equation}\label{equ: Meta epsilon}
\varepsilon (\mathsf{R_I}, L,\xi) \!=\! \sum_{ \theta = L }^{\infty} \!\! \frac{ \theta + 1 - L} { (\theta + 1) !} \bar{\varTheta} (\mathsf{R_I}, \xi) ^\theta  \exp \left(  -\bar{\varTheta} (\mathsf{R_I}, \xi) \right),
\end{equation}
where $\bar{\varTheta}(\mathsf{R_I}, \xi)$ can be either $\bar{\varTheta}^\text{fx} (R_c, \xi)$ or $\bar{\varTheta}^\text{fl} (\mu, \xi) $ given in \eqref{equ: Meta meanTheta}, corresponding to the fixed or flexible IN scheme.

With the IN missing probability, the densities of the interfering BSs in $\Phi_a$, $\Phi_b$, and $\Phi_c$, i.e., $\lambda_a$, $\lambda_b$, and $\lambda_c$, are respectively given by
\begin{align}\label{equ: Meta density of Phi a b c}
\lambda_a& = \varepsilon (\mathsf{R_I}, L,\xi) \lambda \left(1- \frac{1}{\xi} \right), \,\,
\lambda_b = a b  \lambda ,\,\,
\lambda_c = \lambda;
\end{align}
where the coefficients $a $ and $b$ for the two IN schemes are given by
\begin{align} \label{equ: Meta a and b}
a^\text{fx}&= \left\{
\begin{array}{ll}
\varepsilon^\text{fx} (R_c, L, \xi) &  \,  Z<R_c, \\
1 & \,  Z\geq R_c,\\
\end{array}
\right. 
b^\text{fx} = \left\{
\begin{array}{ll}
1 &  \,  Z<R_c, \\
1- {1\over \xi } & \,  Z\geq R_c;\\
\end{array}
\right.\notag \\ 
a^\text{fl}  &= \left\{
\begin{array}{ll}
1-{\frac{1}{\xi} } &  \, \mu<1, \\
1 & \,  \mu\geq 1,\\
\end{array}
\right. \quad
b^\text{fl}  = \left\{
\begin{array}{ll}
1 &  \,  \mu<1, \\
\varepsilon^\text{fl}  (\mu, L, \xi) & \,  \mu\geq 1.\\
\end{array}
\right. 
\end{align}

\subsection{Expressions for the STPs}\label{subsec: Meta Expression for STP}
In this subsection, we will present the expression for the STP of each IN scheme, so as to figure out how the STPs respond to the changes of the design parameters $(\mathsf{R_I}, L,\xi)$, and obtain some system design insights from the perspective of improving the average network performance.

\subsubsection{STP for the Fixed IN Scheme}
As shown in Appendix~\ref{appendix: Meta expression STP}, when file $n$ is requested, the STP for the fixed IN scheme is given by
\begin{align}\label{equ: Meta psn fx}
p_{s,n}^\text{fx} \! = \! \sum_{\theta = 0}^{ L  } \mathbb{P} \left[ \varTheta_I^\text{fx} = \theta \right] \!\!  \int_{0}^{\infty} \!\!\!\! \| \exp \left( \mathbf{Q}_{M-\theta} (R_c, L, \xi) \right) \|_1 f_{Z} (z) \mathrm{d}z,
\end{align}	
where $ \mathbb{P} \left[ \varTheta_I^\text{fx} = \theta \right]$ is given in \eqref{equ: Meta PMF of ThetaI}; $f_{Z} (z) = 2 \pi {\lambda \over \xi} z \exp (-\pi{\lambda \over \xi} z^2)$ is the probability distribution function (PDF) of the serving distance $Z$; $ \exp \left( \mathbf{A} \right) = \sum_{k = 0}^{ \infty} { \mathbf{A}^k \over k! }$ is the matrix exponential; and $\mathbf{Q}_{D} (R_c, L, \xi)$ is a $D \times D$ lower Toeplitz matrix, given by	
\begin{equation}\label{equ: Meta matrixQ fx}
\mathbf{Q}_{D}(R_c, L, \xi) =\left[\begin{array}{cccc}
q_{0} & & & \\
q_{1} & q_{0} & & \\
\vdots & \vdots & \ddots & \\
q_{D-1 } & q_{D-2 }& \cdots & q_{0}
\end{array}\right],
\end{equation}
whose elements, i.e., $q_0$ and $q_m$ for $m \geq 1$ are given by
\begin{align}\label{equ: Meta q_0 fx}
q_0 =& -\pi \lambda \Bigg\{ \varepsilon^\text{fx} (R_c, L, \xi) \left(1- {1\over \xi } \right)  z^{2}  \frac{ \tau ^{ \frac{2} {\alpha}} }{\mathrm{sinc}(\frac{2} {\alpha })} +  a^\text{fx}{ z^2 \over \xi }  F(\tau ) \notag
\\& + b^\text{fx} \left(1- \varepsilon^\text{fx} (R_c, L, \xi) \right) R_c^2 F\left( \tau \left( { z \over R_c} \right)^{ \alpha} \right)   \Bigg\},
\end{align}
\begin{align}\label{equ: Meta q_m fx}
q_m   = &2 \pi \lambda \Bigg\{  \varepsilon^\text{fx} (R_c, L, \xi) \left(1- {1\over \xi } \right)  \frac{\tau^{ \frac{2}{\alpha} } }{\alpha}   z^2 B\left(m-\frac{2}{\alpha}, 1 + \frac{2}{\alpha}\right) \notag
\\& + a^\text{fx}  {1\over \xi }  \tau^m   z^{2} \tilde{F}_{m} ( \tau   )     
+ b^\text{fx} ( 1- \varepsilon^\text{fx} (R_c, L, \xi) )\tau^m   z^{ \alpha m} \notag
\\& \times R_c^{2-\alpha m} \tilde{F}_{m} \left( \tau \left( { z \over R_c} \right)^{ \alpha} \right)   \Bigg\},
\end{align}	
where $\varepsilon^\text{fx} (\mu, L, \xi)$ is shown in \eqref{equ: Meta epsilon}; $\mathrm{sinc}(x) = \frac{\sin \pi x}{\pi x}$; coefficients $a^\text{fx}$ and $b^\text{fx}$ are given by \eqref{equ: Meta a and b}; $B(x,y)$ denotes the Beta function; and $F(x)$ and $ \tilde{F}_{m}(x)$ are respectively given by
\begin{equation}\label{equ: Meta define F(x)}
F(x) = {_2F_{1}}\left(-\frac{2}{\alpha}, 1 ; 1-\frac{2}{\alpha} ;-x \right)-1,
\end{equation}
\begin{equation}\label{equ: Meta Fjk}			
\tilde{F}_{m}(x) =  \frac{1}{\alpha m - 2}  {_2F_{1}} \left(  1 + m, m  -  \frac{2}{\alpha};  m- \frac{2}{\alpha} +  1 ; -x  \right),
\end{equation}
with ${_2F_1}\left( a,b; c;d \right)$ denoting the Gauss hypergeometric function.

Finally, the whole STP for the network with the fixed IN scheme can be obtained by plugging \eqref{equ: Meta psn fx} into \eqref{equ: Meta definition of whole STP}.

\subsubsection{STP for the Flexible IN Scheme}
To present the expression for the STP in this scheme, we first define a notation as follows.
\begin{align}\label{equ: Meta w0 fl}
	w_0   & \triangleq  \varepsilon^\text{fl} (\mu, L, \xi)   \left(1  - {1\over \xi }\right)   \frac{ \tau ^{ \frac{2} {\alpha}} }{\mathrm{sinc}(\frac{2} {\alpha })}   \notag
	\\ & +   a^\text{fl}  \left( 1  -   \varepsilon^\text{fl} (\mu, L, \xi) \right) \mu^2   F  \left(  { \tau \over \mu ^{ \alpha }}  \right)  +  b^\text{fl}  {1\over \xi } F( \tau ) ,
\end{align}	
where $\varepsilon^\text{fl} (\mu, L, \xi)$ is shown in \eqref{equ: Meta epsilon}; $\mathrm{sinc}(x) = \frac{\sin \pi x}{\pi x}$; coefficients $a^\text{fl}$ and $b^\text{fl}$ are given by \eqref{equ: Meta a and b}; and $F(x)$ is given by \eqref{equ: Meta define F(x)}.

We also show in Appendix~\ref{appendix: Meta expression STP} that when file $n$ is requested, the STP for the flexible IN scheme is given by 
\begin{align}\label{equ: Meta psn fl}
p_{s,n}^\text{fl}  =   {1\over \xi } \sum_{\theta = 0}^{ L  } \mathbb{P} \left[ \varTheta_I^\text{fl} = \theta \right] \left\|   \mathbf{W}_{ M - \theta }(\mu, L, \xi)  ^{-1}   \right\|_1 ,
\end{align}		
where $\mathbb{P} \left[ \varTheta_I^\text{fl} = \theta \right]$ is given in \eqref{equ: Meta PMF of ThetaI}; $\|\cdot \|_1$ is the $L_1$ induced matrix norm; $ \mathbf{W}_{ M - \theta }(\mu, L, \xi) \triangleq \left( {1\over \xi } + 2 w_0   \right) \mathbf{I}_{ M - \theta}   -   \tilde{ \mathbf{W} }_{ M - \theta }(\mu, L, \xi) $, where $w_0 $ is given by \eqref{equ: Meta w0 fl}, $\mathbf{I}_{n}$ denotes an $n\times n$ identity matrix, and $\tilde{ \mathbf{W} }_{ D }(\mu, L, \xi)$ is a $D\times D$ lower Toeplitz matrix given by 
\begin{equation}\label{equ: Meta matrixW2 fl}
\tilde{ \mathbf{W} }_{D}(\mu, L, \xi)=\left[\begin{array}{cccc}
w_0 & & & \\
w_{1} & w_0 & & \\
\vdots & \vdots & \ddots & \\
w_{D-1 } & w_{D-2 } & \cdots & w_0
\end{array}\right],
\end{equation}	
the elements of which for $m\geq 1$ are 
\begin{align}\label{equ: Meta wm fl}
w_m \!& =    2 \Bigg\{ \varepsilon^\text{fl} (\mu, L, \xi) \left(1  - {1\over \xi }\right)  \frac{\tau^{ \frac{2}{\alpha } } }{\alpha } B \left(  m - \frac{2}{\alpha }, 1 +  \frac{2}{\alpha}  \right)\notag
\\& + \! a^\text{fl} \! (1 \! - \!  \varepsilon^\text{fl} (\mu, L, \xi)  \! ) \tau^m \mu ^{2-\alpha m} \tilde{F}_{m} \! \left(  { \tau \over \mu^{\alpha} }  \right) \! + \! { b^\text{fl} \over \xi } \tau^m \tilde{F}_{m}(\tau ) \!  \Bigg\},
\end{align}	
where $B(x,y)$ is the Beta function, and $\tilde{F}_{m}(x)$ is given by \eqref{equ: Meta Fjk}.

Finally, the whole STP for the network with the flexible IN scheme can be obtained by plugging \eqref{equ: Meta psn fl} into \eqref{equ: Meta definition of whole STP}.

\subsection{The SIR Meta Distribution}\label{subsec: Meta Approx. for Meta}
The exact expression for the SIR meta distribution of file $n$ can be obtained using the Gil-Pelaez theorem \cite{GIL-PELAEZ1951}, i.e., 
\begin{equation}\label{equ: Meta Gil-Pelaez}
\bar{F}_{P_{s,n}}(x)=\frac{1}{2}+\frac{1}{\pi} \int_{0}^{\infty} \frac{\Im\left(e^{-j t \log x} M_{j t, n } \right)}{t} \mathrm{~d} t, \quad x \in[0,1],
\end{equation}
where $j \triangleq \sqrt{-1}$; $M_{i,n} = \mathbb{E} \left[ P_{s,n}(\tau)^{i} \right]$ denotes the $i$-th moment of the CSTP $P_{s,n}(\tau)$, and then, $M_{jt,n}$ is the corresponding imaginary moment of $P_{s,n}(\tau)$; and $\Im(z)$ gives the imaginary part of $z\in \mathbb{C}$. 

However, as mentioned in Section~\ref{subsec: Meta SIR Model}, the effective channel gain of the desired signal in multi-antenna networks follows Gamma distribution. This will involve the calculation of high-order derivatives of the Laplace transform of the interference, which makes it impossible to derive exact expressions for high-order moments of the $P_{s,n} (\tau)$. Even if $M_{jt,n} $ is given, the calculation of the exact meta distribution using \eqref{equ: Meta Gil-Pelaez} involves the integration of imaginary, which is still tedious. As verified in \cite{Haenggi2016,Deng2017,Cui2018a,Salehi2019} for different network scenarios, the beta distribution provides an excellent approximation for the SIR meta distribution. Specifically, given the first moment $M_{1,n}$ and the second moment $M_{2,n}$ of the CSTP, by matching them with the first and second moments of the beta distribution, we have the following approximation:
\begin{equation}\label{equ: Meta beta approx.}
\bar{F}_{P_{s,n}}(x) \approx 1-I_{x}\left(\frac{M_{1,n} \kappa}{1-M_{1,n}}, \kappa \right), \quad x \in[0,1],
\end{equation}
where $\kappa = \frac{\left(M_{1,n} -M_{2,n} \right) \left(1-M_{1,n} \right)} {\left(M_{2,n} -(M_{1,n}) ^{2}\right)}$; $I_x(a,b)$ is the regularized incomplete beta function with parameters $a,b>0$, i.e., $ I_x(a,b)= \frac{\int_{0}^{x} t^{a-1}(1-t)^{b-1} \mathrm{~d} t}{B(a, b)}$; and $B(a,b) = \int _{0}^{1}t^{a-1}(1-t)^{b-1}\,\mathrm{d}t$ here is the Beta function.

In this work, since the exact expressions for the second moment of $P_{s,n}(\tau)$ cannot be obtained, we resort to a tractable upper bound on $P_{s,n} (\tau)$, denoted by $ P_{s,n}^{u}(\tau)$, to obtain an approximation of the SIR meta distribution. We first present the expression of this upper bound.

As shown in Appendix~\ref{appendix: Meta Upper Bound}, for both IN schemes, an upper bound on $ P_{s,n}(\tau)$ is given by
\begin{align}\label{equ: Meta Psn upper}
	P_{s,n}^{u} (\tau) &= \sum_{\theta = 0}^{L} \mathbb{P} \left[ \varTheta_I = \theta \right]  \sum_{i=1}^{ M-\theta  } (-1)^{i+1} \binom{ M-\theta }{i} \notag
	\\  & \qquad \quad \times \prod_{x \in \Phi_a \cup \Phi_b \cup \Phi_c } {1 \over 1 + i \beta \tau Z^{\alpha} \|x\|^{-\alpha} },
\end{align}
where $\beta = ((M-\theta) !)^{ -1 \over M-\theta }$ and $Z$ is the serving distance of $u_0$ given a realization of $\Phi$. Note that $P_{s,n}^{u} (\tau)$ can be $P_{s,n}^{\text{fx},u}(\tau)$ (for the fixed IN scheme) or $P_{s,n}^{\text{fl},u}(\tau)$ (for the flexible IN scheme), the corresponding $\varTheta_I $ can be $\varTheta_I^\text{fx}$ or $\varTheta_I^\text{fl}$, and $\Phi_i $, $i\in\{a,b,c\} $ can correspondingly be $\Phi_i^{\text{fx}} $ in \eqref{equ: Meta Phi a b c Rc} or $\Phi_i^{\text{fl}} $ in \eqref{equ: Meta Phi a b c}. In the following, we present the expressions for the first and second moments of $ P_{s,n}^{u}(\tau)$ under each IN scheme.

For the fixed IN scheme, as shown in Appendix~\ref{appendix: Meta M1nu M2nu}, by calculating the first and second moments of the upper bound $ P_{s,n}^{\text{fx},u}(\tau)$ given by \eqref{equ: Meta Psn upper}, i.e., $M_{i,n}^{\text{fx},u} = \mathbb{E} \left[ P_{s,n}^{\text{fx},u}(\tau)^i \right]$, $i\in \{1,2 \}$, we can obtain
\begin{align}\label{equ: Meta M1nu fx}
M_{1,n}^{\text{fx},u}  & = \sum_{\theta = 0}^{L} \mathbb{P} \left[ \varTheta_I^\text{fx} = \theta \right]  \sum_{i=1}^{ M-\theta  } (-1)^{i+1} \binom{ M-\theta }{i} \int_{0}^{\infty} f_{Z}(z) \notag
\\  &  \times  \exp \Bigg(   - \pi \lambda   \Bigg( \varepsilon^\text{fx} (R_c, L, \xi) \left(1- {1\over \xi } \right)  z^{2}   \frac{ ( i \beta \tau ) ^{ \frac{2} {\alpha}} }{\mathrm{sinc}(\frac{2} {\alpha })}   \notag
\\ &  + a^\text{fx}{ z^2 \over \xi }  F ( i \beta \tau )  + b^\text{fx} \left(1- \varepsilon^\text{fx} (R_c, L, \xi) \right) R_c^2 \notag
\\ &\times F\left( i \beta \tau \left( { z \over R_c} \right)^{ \alpha} \right)     \Bigg)
\!\Bigg ) \mathrm{d}z,
\end{align}
\begin{align}\label{equ: Meta M2nu fx}
&M_{2,n}^{\text{fx},u}  =   \sum_{\theta_1 = 0}^{ L }  \sum_{\theta_2 = 0}^{ L }  \sum_{i=1}^{M-\theta_1} \sum_{j=1}^{M - \theta_2 } \mathbb{P} \left[ \varTheta_I^\text{fx} = \theta_1 \right]  \mathbb{P} \left[ \varTheta_I^\text{fx} = \theta_2 \right] \notag 
\\&  \times  (-1)^{i+j}  \binom{M-\theta_1}{i} \binom{M - \theta_2 }{j} \int_{0}^{\infty}  f_{Z} (z) \notag 
\\&  \times \exp \left(   -2\pi \sum_{k = a}^c \lambda_k^{\text{fx}}  \mathcal{H}_{ij} (\varOmega_k^{\text{fx}}, \beta_1 \tau z^{\alpha} , \beta_2 \tau z^{\alpha}  ) \right ) \mathrm{d}z,
\end{align}
where $\beta = ((M-\theta) !)^{ -1 \over M-\theta }$, and $\beta_m= ( (M-\theta_m) !)^{-1\over M-\theta_m }$, $m \in \{1,2\}$; $\mathrm{sinc}(x) = \frac{\sin \pi x}{\pi x}$; coefficients $a^{\text{fx}} $ and $b^{\text{fx}} $ are given in \eqref{equ: Meta a and b}; $f_{Z} (z) = 2 \pi {\lambda \over \xi} z \exp (-\pi{\lambda \over \xi} z^2)$ is the PDF of the serving distance $Z$; $\varOmega_k^{\text{fx}}$ and $\lambda_k^{\text{fx}}$, $k \in \{ a,b,c \}$ are given in \eqref{equ: Meta Omega abc Rc} and \eqref{equ: Meta density of Phi a b c}, respectively; $\mathcal{H}_{ij} (\varOmega,x,y)$ is given by
\begin{equation}\label{equ: Meta Hij}
\mathcal{H}_{ij} (\varOmega,x,y) \triangleq \int_{\varOmega} \left( 1-    { 1 \over  \left( 1 + i x v^{-\alpha} \right)\left( 1 + j y v^{-\alpha} \right) }  \right) v \mathrm{d} v,
\end{equation}
where $\varOmega$ denotes the integration limits; we note that $\mathcal{H}_{ij} (\varOmega,x,y) = \mathcal{H}_{ji} (\varOmega,y,x)$, which can save computational time significantly.

Similarly, for the flexible IN scheme, as shown in Appendix~\ref{appendix: Meta M1nu M2nu}, by calculating the first and second moments of the upper bound $ P_{s,n}^{\text{fl},u}(\tau)$ given by \eqref{equ: Meta Psn upper}, i.e., $M_{i,n}^{\text{fl},u} = \mathbb{E} \left[ P_{s,n}^{\text{fl},u}(\tau)^i \right]$, $i\in \{1,2 \}$, we can obtain
\begin{align}\label{equ: Meta M1nu fl}
&M_{1,n}^{\text{fl},u}  = \sum_{\theta = 0}^{L} \mathbb{P} \left[ \varTheta_I^\text{fl} = \theta \right]  \sum_{i=1}^{ M-\theta  } (-1)^{i+1} \binom{ M-\theta }{i}   \notag
\\&  \times \Big( 1+ \xi \Big(  \varepsilon^\text{fl} (\mu, L, \xi)  (1  - {1\over \xi } )   \frac{ ( i \beta \tau ) ^{ \frac{2} {\alpha}} }{\mathrm{sinc}(\frac{2} {\alpha })}   \notag
\\ &  +   a^\text{fl}  \left( 1  -   \varepsilon^\text{fl} (\mu, L, \xi) \right) \mu^2   F   (  { i \beta \tau \over \mu ^{ \alpha }}   )  + { b^\text{fl} \over \xi } F( i \beta \tau )  \Big)
\!\Big ) ^{-1},
\end{align}                          
and the expression for $M_{2,n}^{\text{fl},u}$ is similar to that for $M_{2,n}^{\text{fx},u}$, and can be obtained by substituting all the superscript ``fx'' in \eqref{equ: Meta M2nu fx} with ``fl''.

Note that the aforementioned expressions for $M_{1,n}^{u} $ and $M_{2,n}^{u} $ involve a large number of summation and integration operations, which may make the numerical computation time-consuming when the values of $M$ and $L$ are large. However, the approximation framework considered in the paper is still of great importance, since it is impossible to obtain an explicit expression for the second moment if we focus on the exact expression of the CSTP $P_{s,n} (\tau)$.

After obtaining the expressions for $M_{1,n}^{u} $ and $M_{2,n}^{u} $, we substitute them into \eqref{equ: Meta beta approx.}. Then, the corresponding left-hand side of \eqref{equ: Meta beta approx.} is rewritten as $ \bar{F}_{P_{s,n}^{u} }(x)$. In this case, \eqref{equ: Meta beta approx.} gives a beta approximation of $ \bar{F}_{P_{s,n}^{u} }(x)$. In the sequel, we will use this beta approximation of $ \bar{F}_{P_{s,n}^{u} }(x)$ as an approximation of the exact SIR meta distribution to analyze the link reliability distribution of the network. The tightness of this approximation will be verified via simulation in Section~\ref{section: Meta Numerical}. The overall SIR meta distribution for both IN schemes then can be approximated by 
\begin{equation}\label{equ: Meta barF Pus}
\bar{F}_{P_{s}}(x) \approx \sum_{ n \in \mathcal{N}_c} a_n  \bar{F}_{P_{s,n}^{u} } (x).
\end{equation} 

Together with the STP, the SIR meta distribution forms a holistic analysis framework to reflect the performance from the perspectives of not only the overall network but also the individual links.

\begin{figure*}[!t]	
	\centering
	\subfloat[The STP and the upper bound]{\includegraphics[width=5.7cm]{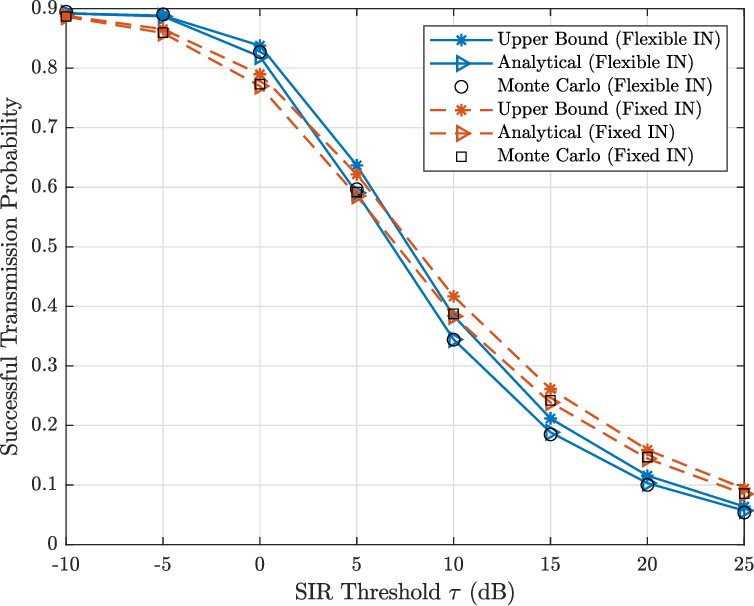} \label{fig: Meta sub Verify STP} }\,\,
	\subfloat[The approximation of the variance] {\includegraphics[width=5.7cm]{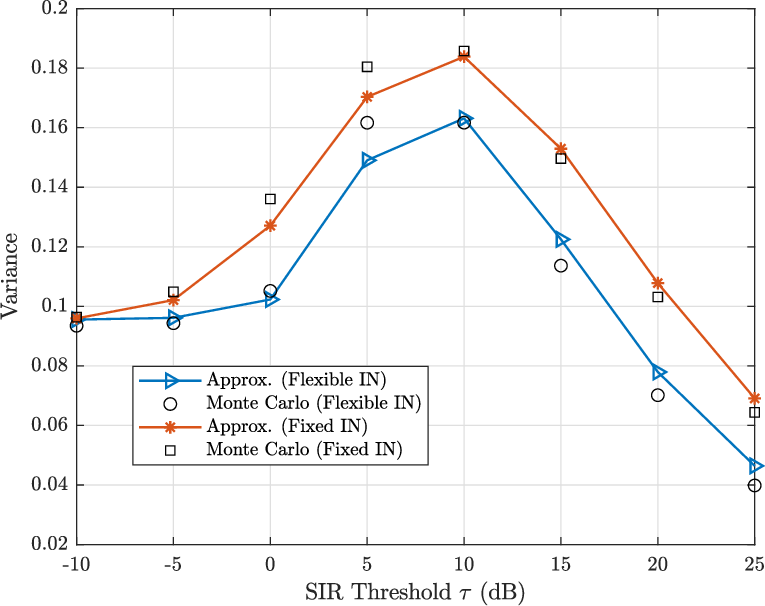} \label{fig: Meta sub Verify Var}}\,\,
	\subfloat[The approximation of the meta distribution] {\includegraphics[width=5.7cm]{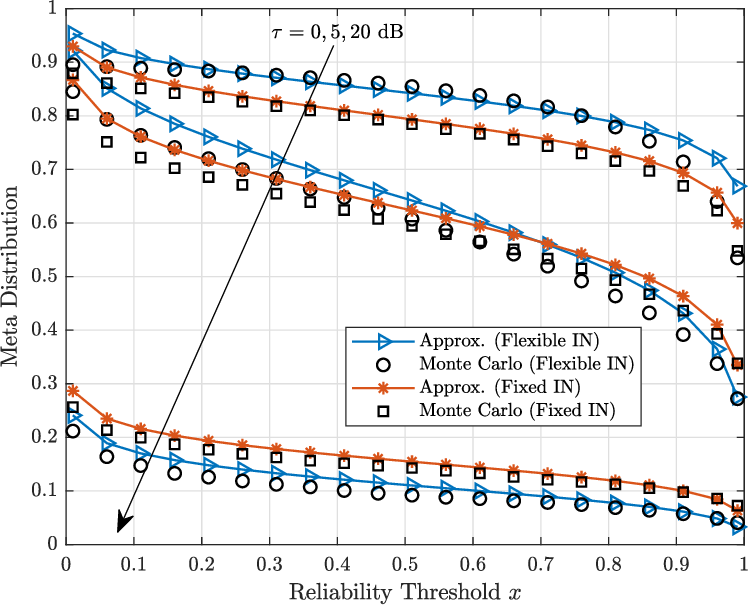} \label{fig: Meta sub Verify Meta}}
	\caption{Verification of the analytical results of the STP, the upper bound on the STP, the approximation of the variance, and the approximation of the SIR meta distribution for the two IN schemes. }
	\label{fig: Meta Verify STP Var Meta}
\end{figure*}

\section{Numerical Results}\label{section: Meta Numerical}	
In this section, we present numerical results to verify the derived analytical expressions and give some design insights.	
Unless otherwise stated, the default simulation parameter settings are as follows: $M = 8$, $L = 2$, $\mu = 0.8$, $P = 46 \, \mathrm{dBm}$, $\alpha = 4$, $\lambda = 1 \times 10^{-4} \, \mathrm{m}^{-2} $, $\lambda_{u} =  8 \times 10^{-4} \, \mathrm{m}^{-2} $, $N=100$, $C = 40$, $\xi = 1.75$. The file popularity is assumed to follow the Zipf distribution, i.e., $a_n = \frac{n^{-\gamma_z}}{\sum_{n\in \mathcal{N}} n^{-\gamma_z}}, \,\,n \in \mathcal{N}$, where $\gamma_z= 0.8$ is the Zipf exponent. 

To fairly compare the performance of the fixed IN scheme and the flexible IN scheme, we use the criterion that the average numbers of IN requests received by a BS under these two schemes are equal, which means the equality $\bar{\varTheta}^\text{fx} (R_c, \xi) =\bar{\varTheta}^\text{fl} (\mu, \xi)  $ holds. By solving this equality via the ``fsolve'' function in MATLAB, given a value of the IN coefficient $\mu$ (resp. the IN range $R_c$) for the flexible IN scheme (resp. the fixed IN scheme), the corresponding value of $R_c$ (resp. $\mu$) can be obtained. For instance, with the above default parameter settings, the corresponding value of $R_c$ is $R_c = 52.71\, \mathrm{m}$.

Moreover, we further consider the variance of the CSTP to evaluate the fairness of individual links among users. By leveraging $M_{1,n}^{u}$ and $M_{2,n}^{u}$ given in Section~\ref{subsec: Meta Approx. for Meta}, we can obtain an approximated variance of the CSTP when file $n$ is requested, which is given by 
\begin{equation}\label{equ: Meta Vn}
V_n \triangleq M_{2,n}^{u} - (M_{1,n}^{u})^2.
\end{equation}
Using the total probability theorem, the variance of the whole CSTP is given by
\begin{align}
V = \sum_{ n \in \mathcal{N}_c} a_n V_n. \label{equ: Meta V}
\end{align}

\subsection{Verification of the Analytical Results}

This subsection compares the analytical results obtained in this paper with Monte Carlo simulations. For the Monte Carlo simulations, we first generate $1\times 10^4$ realizations of $\Phi$ and $\Phi_u$. For each realization of the PPPs, the files cached at each BS and the file requested by each user are respectively determined according to our FUDC policy and the file popularity distribution, so that the user association and the IN request for each BS are determined. Next, for each realization of the $\Phi$ and $\Phi_u$, we further generate $5\times 10^3$ realizations of the Rayleigh fading channel. By averaging the results for these $5\times 10^3$ realizations, we obtain the CSTP for one realization of $\Phi$ and $\Phi_u$. Finally, we obtain $1\times 10^4$ different CSTPs, based on which, the STP, the variance of the CSTP, and the meta distribution are obtained.

Fig.~\ref{fig: Meta Verify STP Var Meta} verifies the analytical results under the fixed IN scheme and flexible IN scheme. Specifically, for different SIR thresholds $\tau$, Fig.~\ref{fig: Meta Verify STP Var Meta}\subref{fig: Meta sub Verify STP} shows the results of the analytical expressions for the STPs $p_{s}^\text{fx}$ and $p_{s}^\text{fl}$, the upper bounds on STPs, i.e., $M_{1}^{\text{fx},u} $ and $M_{1}^{\text{fl},u} $, and the Monte Carlo simulation of the STPs for the two IN schemes. Here, the upper bound on STP for each IN scheme is defined by $M_{1}^{u} \triangleq \sum_{ n \in \mathcal{N}_c} a_n M_{1,n}^{u}$, which is exactly the first moment of the upper bound of the CSTP. As we can observe, for both schemes, the analytical results match the Monte Carlo results well and the upper bounds $M_{1}^{\text{fx},u} $ and $M_{1}^{\text{fl},u} $ are respectively tight to $p_{s}^\text{fx}$ and $p_{s}^\text{fl}$. Fig.~\ref{fig: Meta Verify STP Var Meta}\subref{fig: Meta sub Verify Var} shows that the analytical expressions for the variances $V$ given in \eqref{equ: Meta V} also match the Monte Carlo results. 

For both IN schemes, the accuracy of the upper bounds $M_{1}^{\text{fx},u} $ and $M_{1}^{\text{fl},u} $ as well as the variances $V^\text{fx}$ and $V^\text{fl}$ allows us to use the first and second moments of the upper bound on CSTPs to approximate the moments of the original CSTPs, and further obtain the corresponding approximations of the SIR meta distributions using \eqref{equ: Meta beta approx.} and \eqref{equ: Meta barF Pus}, as illustrated in Section~\ref{subsec: Meta Approx. for Meta}. Fig.~\ref{fig: Meta Verify STP Var Meta}\subref{fig: Meta sub Verify Meta} verifies this idea. The Monte Carlo results in Fig.~\ref{fig: Meta Verify STP Var Meta}\subref{fig: Meta sub Verify Meta} are from the original CSTPs, whereas the analytical results are from \eqref{equ: Meta beta approx.} and \eqref{equ: Meta barF Pus}. From the figure, we observe that the expressions for $\bar{F}_{P_{s}}(x)$ match the Monte Carlo results well, and can effectively reflect the change tendency of the meta distributions for both schemes. Therefore, in the sequel, for each IN scheme, we use the approximation $\bar{F}_{P_{s}}(x)$ in \eqref{equ: Meta barF Pus} to depict the original SIR meta distribution and for simplicity, we refer to it as the meta distribution of the network.

From Fig.~\ref{fig: Meta Verify STP Var Meta}, we see that the variance and meta distribution can provide more information about the individual links than the STP. For instance, in Fig.~\ref{fig: Meta Verify STP Var Meta}\subref{fig: Meta sub Verify STP}, the analytical results of STPs for the two IN schemes are nearly equal when $\tau = 5 \, \mathrm{dB}$. However, Fig.~\ref{fig: Meta Verify STP Var Meta}\subref{fig: Meta sub Verify Var} shows that at this SIR threshold, the variance of the flexible IN scheme is smaller than that of the fixed IN scheme, which means the former scheme brings more fairness among different individual links. Furthermore, from Fig.~\ref{fig: Meta Verify STP Var Meta}\subref{fig: Meta sub Verify Meta}, we see that at this SIR threshold, $46.3\%$ of users achieve link reliability of at least $0.9$ with the fixed IN scheme, whereas $43.1\%$ of users with the flexible IN scheme can reach such reliability. Therefore, although the STPs are nearly equal, the fixed IN scheme has more high-reliability links than the flexible one. This highlights the importance of this paper in further investigating the variance and meta distribution of the network.

In the following, we will further investigate the effects of the IN range $\mathsf{R_I}$, the maximum IN DoF $L$, and the file diversity gain $\xi$ on the STP, the variance, and the meta distribution. For the SIR threshold, we consider a low SIR threshold region and a high SIR threshold region, whose typical values are chosen as $\tau = -6 \, \mathrm{dB}$ and $\tau = 20 \, \mathrm{dB}$, respectively. Moreover, we focus on the meta distribution where the link reliability threshold is given, whose default value is denoted by $x_0$. We set $x_0=0.9$ and study the performance of $\bar{F}_{P_{s}}(x_0)$, which means the fraction of users achieving the link reliability of at least $0.9$ is of interest.

\begin{figure*}[!t]	
	\centering
	\subfloat[The successful transmission probability] {\includegraphics[width=5.7cm]{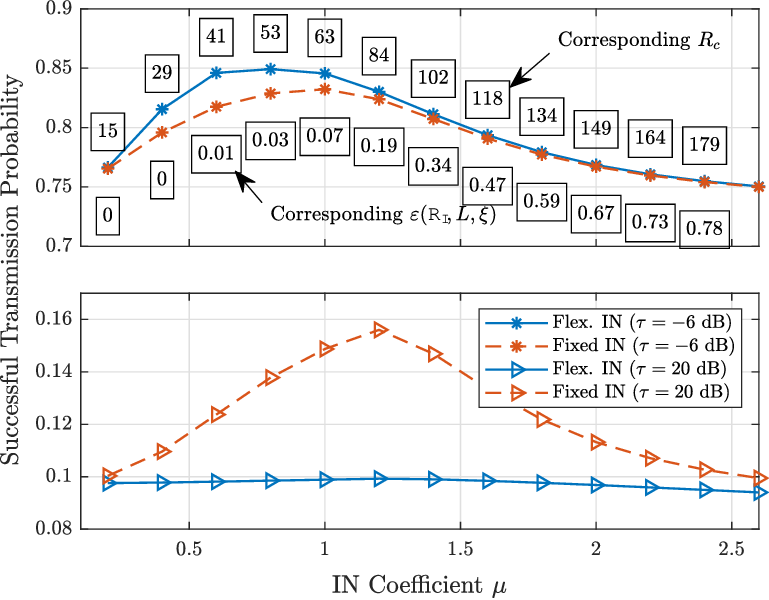} \label{fig: Meta sub CompareSTP Epsilon} }
	\subfloat[The variance] {\includegraphics[width=5.7cm]{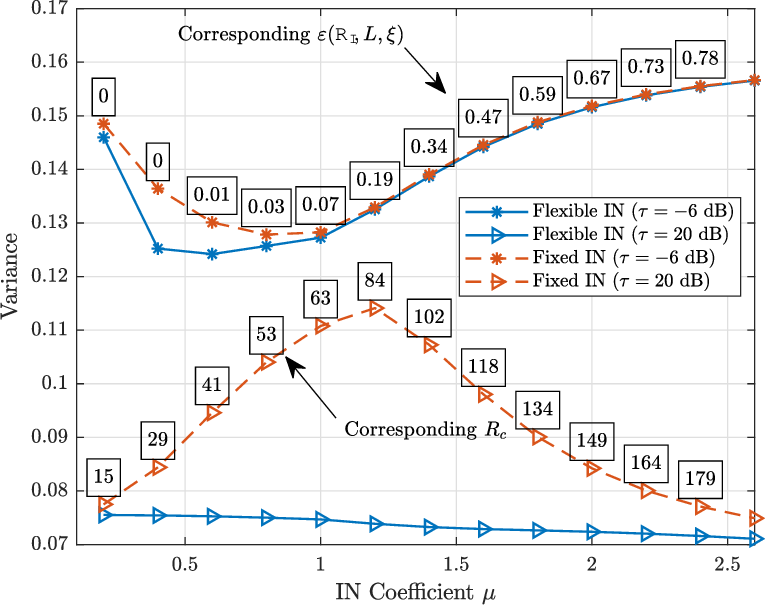} \label{fig: Meta sub CompareVar Epsilon}}
	\subfloat[The meta distribution] {\includegraphics[width=5.7cm]{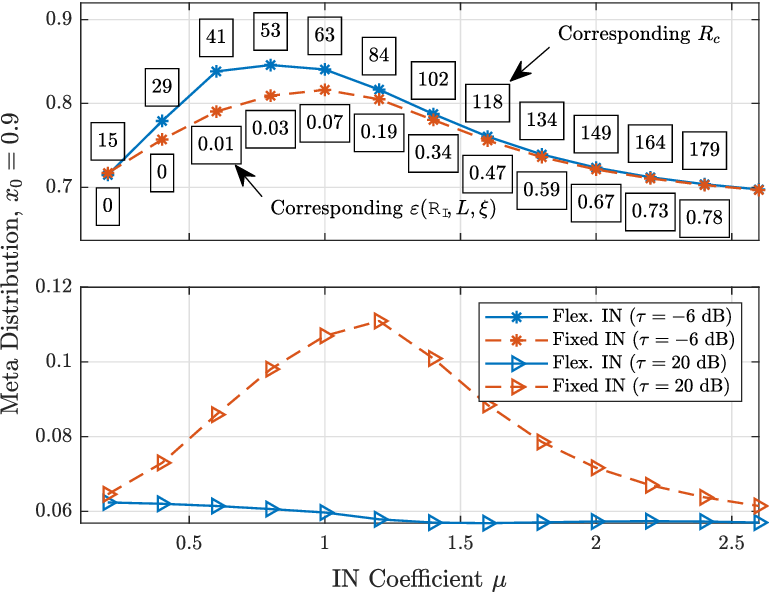} \label{fig: Meta sub CompareMeta Epsilon}}
	\caption{The STP, variance, and meta distribution ($x_0 = 0.9$) versus the IN range $\mathsf{R_I}$ for different SIR thresholds $\tau$. }
	\label{fig: Meta Compare Epsilon}
\end{figure*}

\subsection{Effects of the IN Range $\mathsf{R_I}$ }\label{subsec: Meta Numerical Effect of epsilon}	
In this subsection, we focus on analyzing the effects of the IN range $\mathsf{R_I}$ on the aforementioned three metrics. Note that for the fixed IN scheme, $ \mathsf{R_I}$ represents the IN range $R_c$; whereas for the flexible IN scheme, $ \mathsf{R_I}$ represents the IN coefficient $\mu$. As mentioned before, we choose the value of $R_c$ by making $\bar{\varTheta}^\text{fx} (R_c, \xi) = \bar{\varTheta}^\text{fl} (\mu, \xi)$ hold for each given $\mu$. 

Fig.~\ref{fig: Meta Compare Epsilon} shows the STP $p_{s}$, the variance $V$, and the meta distribution $\bar{F}_{P_{s}}(x_0)$ as functions of IN range $ \mathsf{R_I}$ for different SIR thresholds. $\mu$ is selected as the variable of $x$-axis. For a given $\mu$, the corresponding values of $R_c$ for the fixed IN scheme and $\varepsilon (\mathsf{R_I}, L,\xi)$ are marked within the rectangles shown in the three sub-figures. From Fig.~\ref{fig: Meta Compare Epsilon}\subref{fig: Meta sub CompareSTP Epsilon} and Fig~\ref{fig: Meta Compare Epsilon}\subref{fig: Meta sub CompareMeta Epsilon}, we can see that with the change of $ \mathsf{R_I}$, regarding $p_s$ and $\bar{F}_{P_{s}}(x_0)$, the flexible IN scheme outperforms the fixed one in the low SIR threshold region, whereas the fixed IN scheme performs better in the high SIR threshold region. Fig.~\ref{fig: Meta Compare Epsilon}\subref{fig: Meta sub CompareVar Epsilon} shows that the variance with the flexible IN scheme is always smaller than that with the fixed IN scheme, which means the flexible IN scheme makes the link quality of different users fairer.

Moreover, from Fig.~\ref{fig: Meta Compare Epsilon}\subref{fig: Meta sub CompareSTP Epsilon}, we observe that the STPs for both IN schemes first increase and then decrease with $ \mathsf{R_I}$, which indicates that we should carefully select the values of $\mathsf{R_I}$ to maximize the STP (though the effect of adjusting $\mu$ on the STP under the flexible IN scheme in the high SIR threshold region is slight). 

Regarding the variance, from Fig.~\ref{fig: Meta Compare Epsilon}\subref{fig: Meta sub CompareVar Epsilon}, we observe that there also exists an optimal $\mathsf{R_I}$ for minimizing the variance. Note that the optimal values of $\mathsf{R_I}$ for maximizing the STP and minimizing the variance are different. This indicates that improving the whole STP of the network may compromise the fairness of link reliability among different users, while focusing on improving the link fairness among different users may not achieve optimal overall performance. This reflects the importance of link reliability distribution analysis in network design.

Fig.~\ref{fig: Meta Compare Epsilon}\subref{fig: Meta sub CompareMeta Epsilon} shows that $\bar{F}_{P_{s}}(x_0)$ follows the similar tendency to $p_s$ when $\mathsf{R_I}$ changes. Nevertheless, we can still obtain more detailed information on the link reliability distribution in the network. For example, when $\varepsilon (\mathsf{R_I}, L,\xi) = 0.07$ (i.e., $\mu = 1$), the fractions of users achieving the link reliability of at least $0.9$ are respectively $81\%$ and $84\%$ with the fixed IN scheme and the flexible IN scheme when $\tau= -6 \, \mathrm{dB} $; whereas those fractions are nearly $11\%$ and $6\%$ when $\tau= 20 \, \mathrm{dB}$.

\begin{figure*}[!t]	
	\centering
	\subfloat[The successful transmission probability] {\includegraphics[width=5.7cm]{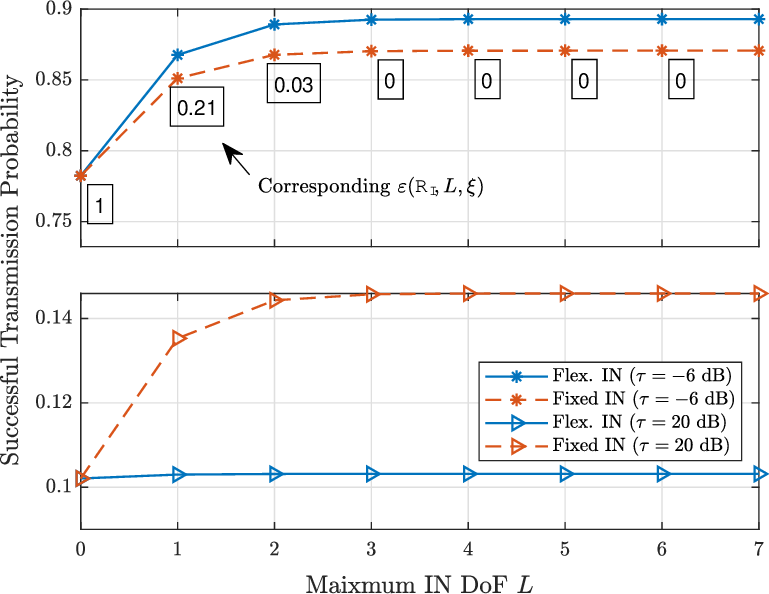} \label{fig: Meta sub CompareSTP OO} }
	\subfloat[The variance] {\includegraphics[width=5.7cm]{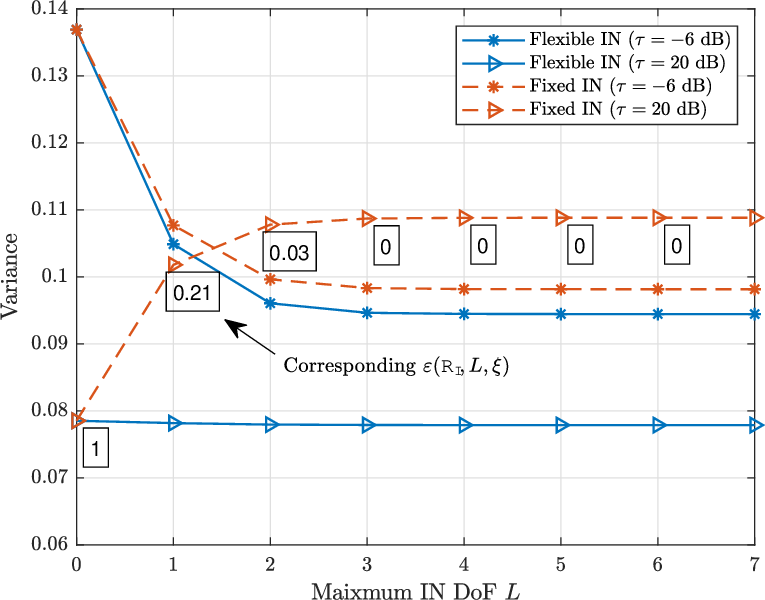} \label{fig: Meta sub CompareVar OO}}
	\subfloat[The meta distribution] {\includegraphics[width=5.7cm]{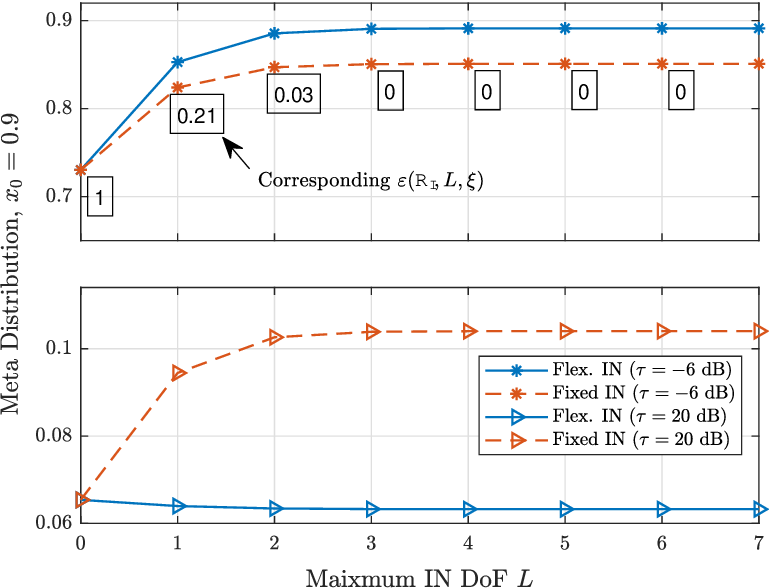} \label{fig: Meta sub CompareMeta OO}}
	\caption{The STP, variance, and meta distribution ($x_0 = 0.9$) versus the maximum IN DoF $L$ for different SIR thresholds $\tau$. }
	\label{fig: Meta Compare OO}
\end{figure*}

\subsection{Effects of the Maximum IN DoF $L$}\label{subsec: Meta Numerical Effect of L}	

This subsection focuses on analyzing the effects of the maximum IN DoF $L$ on the STP, the variance, and the meta distribution. Fig.~\ref{fig: Meta Compare OO} depicts the relations between these three metrics and $L$ in the low SIR threshold region and the high SIR threshold region. Given a value of $L$, the corresponding IN missing probability is marked within the rectangle shown in the sub-figures (recalling that we set the same value of $\varepsilon (\mathsf{R_I}, L,\xi)$ for the two IN schemes by adjusting the value of $R_c$ given a $\mu$). 

From Fig.~\ref{fig: Meta Compare OO}\subref{fig: Meta sub CompareSTP OO}, we observe that the STPs for both schemes first increase and then remain unchanged with $L$. This is because when $L$ is small, the antenna resource allocated to IN is not enough to satisfy all the IN requests received at each BS. Therefore, increasing $L$ will make more IN requests be satisfied, and hence the STP improved. When $L$ is sufficiently large, almost all the IN requests are satisfied, e.g., when $L=2$, the IN missing probability is $\varepsilon (\mathsf{R_I}, L,\xi) = 0.03$, which means averagely $97\%$ of the IN requests can be satisfied. In this case, keeping increasing $L$ has little effect on improving the STP.

From the perspective of the variance, we can observe from Fig.~\ref{fig: Meta Compare OO}\subref{fig: Meta sub CompareVar OO} that under the flexible IN scheme, increasing $L$ will decrease the variance. As we expect, allocating more antenna resources for IN with the flexible scheme will always be beneficial to link fairness. However, for the fixed IN scheme, only in the low SIR threshold region, increasing $L$ can decrease the variance; in the high SIR threshold region, increasing $L$ will increase the variance. Note that allocating more antennas for IN decreases the available spatial diversity for boosting the desired signal. In the low SIR threshold region, the SIR threshold is easy to achieve. For both IN schemes, increasing the DoF for IN will improve the link quality of the poor-link-quality users to achieve the SIR threshold, since more IN requests can be satisfied, and hence, the variance decreases. However, in the high SIR threshold region, the fixed IN scheme does not bring enough gain for the poor-link-quality users to achieve the SIR threshold, but makes the link quality of users who have already reached the SIR threshold better, which is detrimental to the link fairness. 

Fig.~\ref{fig: Meta Compare OO}\subref{fig: Meta sub CompareMeta OO} depicts the relation between $\bar{F}_{P_{s}}(x_0)$ and $L$. We can observe that in the low SIR threshold region, compared with the fixed scheme, the flexible IN scheme leads to a higher fraction of users achieving the link reliability of at least $0.9$. Increasing $L$ will increase the fraction, and when $L$ is large enough, $\bar{F}_{P_{s}}(x_0)$ almost keeps unchanged. Nevertheless, in the high SIR threshold region, the fixed scheme leads to better performance. Moreover, in this SIR threshold region, under the flexible scheme, increasing $L$ is harmful to $\bar{F}_{P_{s}}(x_0)$, since more antennas are allocated for IN to improving the received SIR of the poor-link-quality users, leading to a decrease in the fraction of high-link-quality users.

\begin{figure*}[!t]	
	\centering
	\subfloat[The successful transmission probability] {\includegraphics[width=5.7cm]{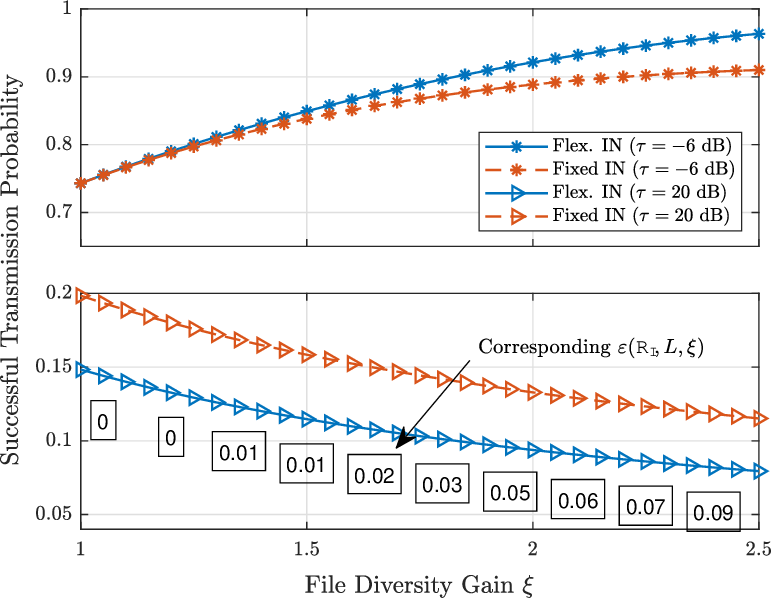} \label{fig: Meta sub CompareSTP Xi} }
	\subfloat[The variance] {\includegraphics[width=5.7cm]{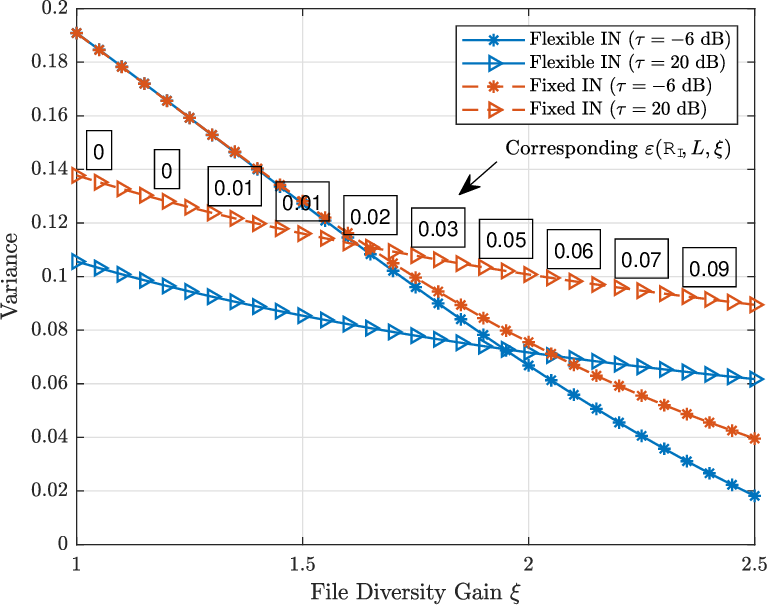} \label{fig: Meta sub CompareVar Xi}}
	\subfloat[The variance] {\includegraphics[width=5.7cm]{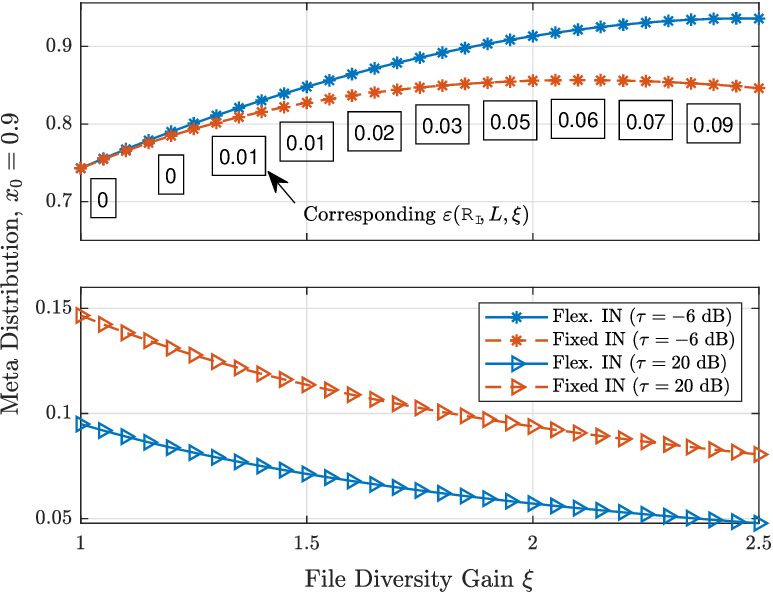} \label{fig: Meta sub CompareMeta Xi}}
	\caption{The STP, variance, and meta distribution ($x_0 = 0.9$) versus the file diversity gain $\xi$ for different SIR thresholds $\tau$. }
	\label{fig: Meta Compare Xi}
\end{figure*}
\subsection{Effects of the File Diversity Gain $\xi$}\label{subsec: Meta Numerical Effect of Xi}	

This subsection focuses on analyzing the effects of the file diversity gain $\xi$ on the STP, the variance, and the meta distribution. Fig.~\ref{fig: Meta Compare Xi} shows these three metrics as functions of the file diversity gain $\xi$ in different SIR threshold regions. Given a value of $\xi$, the corresponding IN missing probability is marked within the rectangle shown in the sub-figures. 

From Fig.~\ref{fig: Meta Compare Xi}\subref{fig: Meta sub CompareSTP Xi}, we observe that regarding the STP, in the low SIR threshold region, the performance of the two IN schemes is nearly the same when $\xi$ is small, and the flexible IN scheme outperforms the fixed one when $\xi$ is large. For both schemes, the STP is an increasing function of $\xi$. In the high SIR threshold region, the fixed one always performs better than the flexible one, and the STPs decrease with $\xi$. This indicates that in the low SIR threshold region, caching more files (corresponding to the UDC policy, $\xi=2.5$) in the network is beneficial to the STP, since the SIR threshold is easy to reach, larger $\xi$ means more files can be found in the network, which is better for improving the STP; whereas in the high SIR threshold region, caching the most popular files is a better choice (exactly the MPC policy, $\xi=1$), since the SIR threshold is hard to reach, ensuring the successful transmission of the most popular files is better. 

Moreover, as can be observed from Fig.~\ref{fig: Meta Compare Xi}\subref{fig: Meta sub CompareVar Xi}, increasing the file diversity gain in the low SIR threshold region is not only beneficial to the STP but the user link fairness (i.e., decreasing the variance). However, in the high SIR threshold region, even though selecting $\xi=1$ contributes to the maximum STP, it also makes the link reliability of different users unfair. Furthermore, regarding the user link fairness, compared with the high SIR threshold region, the benefit brought by increasing $\xi$ is more significant in the low SIR threshold region. For example, when $\tau= -6 \, \mathrm{dB} $, the percentage of the variance decreased by increasing the file diversity gain from $\xi=1$ to $\xi = 2.5$ with the flexible IN scheme is $90.5\%$; whereas when $\tau= 20 \, \mathrm{dB} $, that percentage is $41.5\%$.

Fig.~\ref{fig: Meta Compare Xi}\subref{fig: Meta sub CompareMeta Xi} shows that in the low SIR threshold region, $\bar{F}_{P_{s}}(x_0)$ is an increasing function of $\xi$ for both IN schemes. The two schemes perform almost the same when $\xi$ is small, whereas the flexible IN scheme performs better than the fixed one when $\xi$ is large. Moreover, in the high SIR threshold region, $\bar{F}_{P_{s}}(x_0)$ decreases with $\xi$ for both schemes, and the fixed IN scheme performs better than the flexible one.

\begin{figure}[!t]
	\centering
	\includegraphics[scale = 0.5]{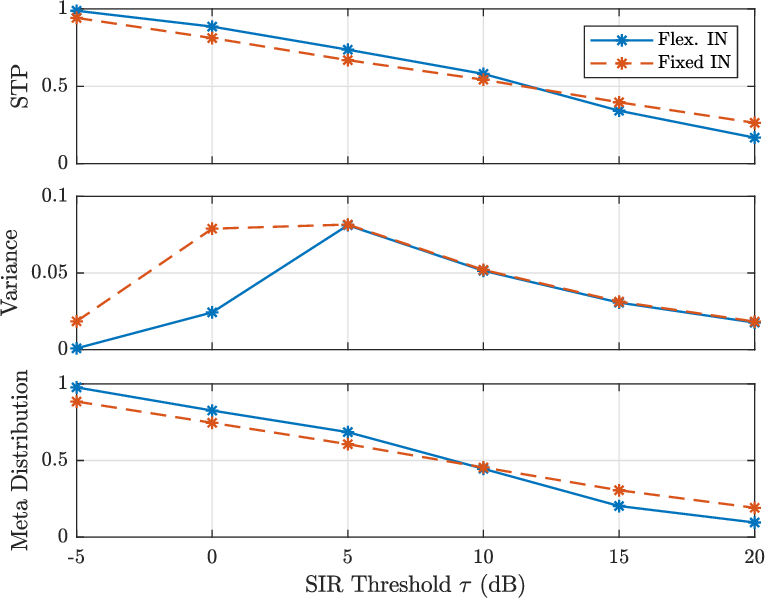}
	\caption{Values of performance metrics under different SIR thresholds $\tau$ when optimizing the STP, the variance and the meta distribution ($x_0 = 0.9$).}
	\label{fig: Meta FigS_Opt_SingleMetric}
\end{figure}	

\subsection{Performance Optimization}\label{subsec: Meta performance optimization}
In this subsection, we endeavor to investigate several optimization problems with the metrics considered in this paper to obtain some system design guidelines quantitatively. The optimization problem can be formulated as:
\begin{problem}\label{problem: Meta general Prob}
	\begin{subequations}
		\begin{align}
		( \mathsf{R_I} ^\star, L ^\star, \xi ^\star)& = \arg \underset{  \mathsf{R_I},\, L,  \, \xi}{\max} \,\, S \label{equ: Meta general Metric}  \\
		\text{s.t.} 
		\quad  \mathsf{R_I} &\geq 0, \label{equ: Meta general Prob Constraint RI}\\
		\quad 0 \leq  L  &\leq M-1, \label{equ: Meta general Prob Constraint L}\\
		\quad  1 \leq  \xi & \leq \frac{N}{C}, \label{equ: Meta general Prob Constraint xi}
		\end{align}		
	\end{subequations}
\end{problem}
where, for both fixed and flexible IN schemes, $S$ can be either $p_s$ given in Section~\ref{subsec: Meta Expression for STP}, or $\bar{F}_{P_{s}}(x)$ given in \eqref{equ: Meta barF Pus}, or $1/V$ with $V$ given in \eqref{equ: Meta V}, or a weighted summation of $p_s$ and $\bar{F}_{P_{s}}(x_0)$, i.e., $\eta p_s + (1 - \eta)  \bar{F}_{P_{s}}(x_0)$, where $\eta \in [0,1]$ is a weight coefficient to balance the priority of maximizing the STP or maximizing the fraction of users achieving a given link reliability;\footnote{One can choose different combinations from $p_s$, $\bar{F}_{P_{s}}(x)$, and $1/V$ to form different weighted objective functions for different purposes. Here, as an example, our goal is to maximize the average STP of the networks as well as to improve the fraction of users achieving a given link reliability $x_0$.} and $( \mathsf{R_I} ^\star, L ^\star, \xi ^\star)$ are the optimal values for the three parameters for each problem. Note that for the weighted objective function, a larger value of $\eta$ means we pay more attention to increasing the STP.

\begin{figure*}[!t]	
	\centering
	\subfloat[Maximize the STP] {\includegraphics[width=5.7cm]{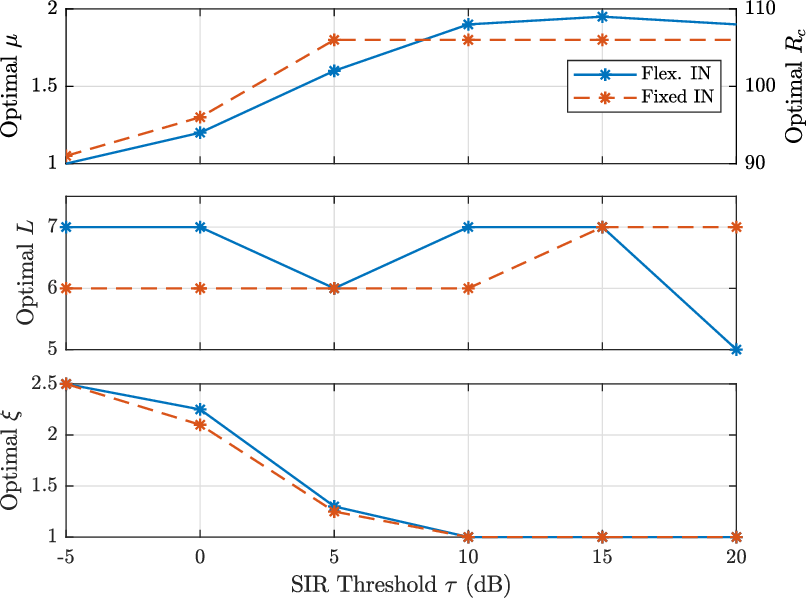} \label{fig: Meta sub FigS_Opt_SingleMaxqPara} }\,\,
	\subfloat[Minimize the variance] {\includegraphics[width=5.7cm]{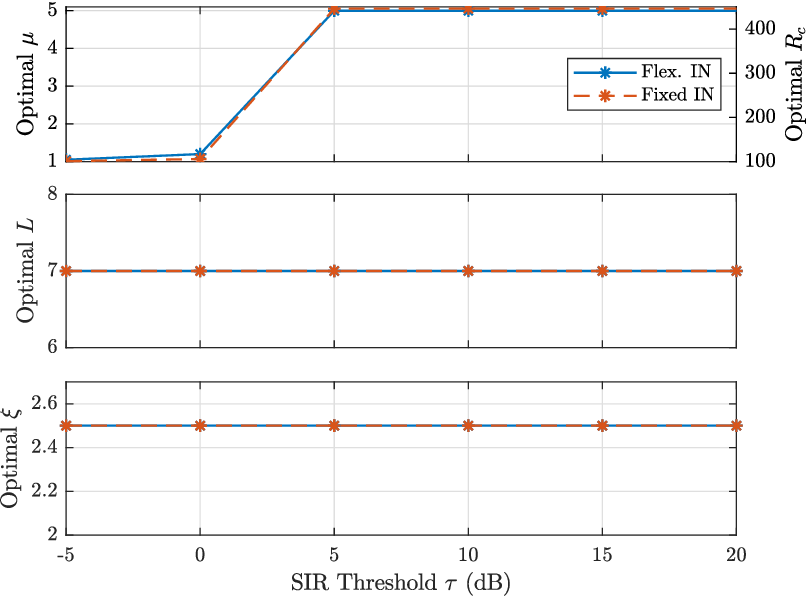} \label{fig: Meta sub FigS_Opt_SingleMinVarPara}}\,\,
	\subfloat[Maximize the meta distribution ($x_0=0.9$)] {\includegraphics[width=5.7cm]{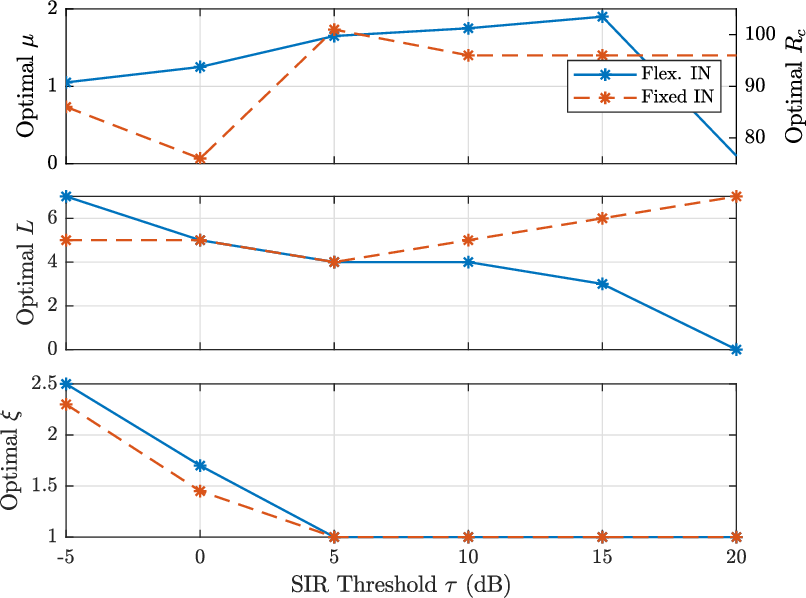} \label{fig: Meta sub FigS_Opt_SingleMaxMetaPara}}
	\caption{The optimal design parameters with different SIR thresholds $\tau$ for optimizing the STP, the variance and the meta distribution.}
	\label{fig: Meta SingleOptimizationDesignPara}
\end{figure*}	

Problem~\ref{problem: Meta general Prob} can be iteratively solved by using the coordinate descent method \cite{DimitriP.Bertsekas2016}. Specifically, in each iteration, we successively search for the optimal value of each variable (e.g., $\mathsf{R_I}$) numerically with the remaining two variables (e.g., $L$ and $\xi$) fixed. Then, by repeating this procedure, an stationary point for this optimization problem can be obtained. However, due to the complexity of the expressions for the objective function, we can only reach an approximated stationary point where the precision of the variables $\mathsf{R_I}$, $ L $, and $\xi$ are $\epsilon_\mathsf{R_I} $, $\epsilon_L$, and $\epsilon_\xi$, whose values should be carefully selected, so that the computation time cost is acceptable. 

\begin{figure}[!t]
	\centering
	\includegraphics[scale = 0.5]{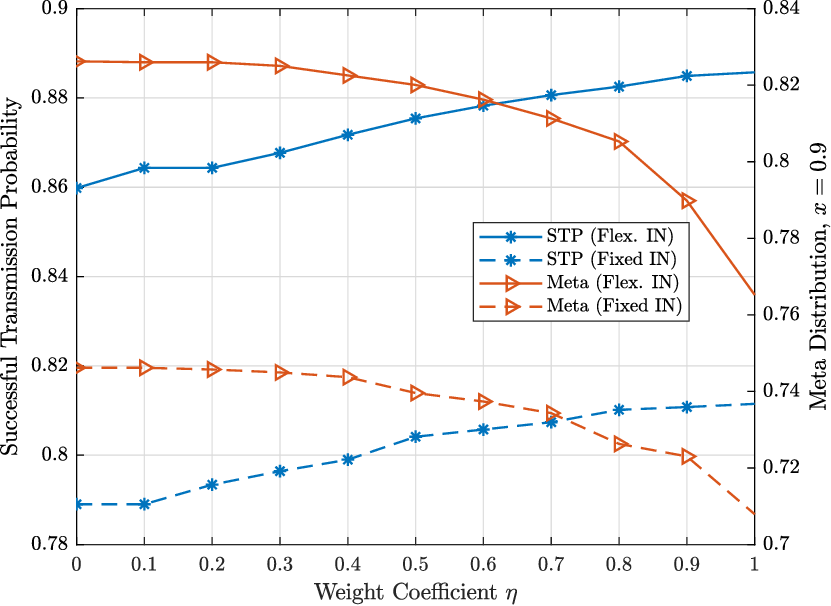}
	\caption{Impact of the weight coefficient $\eta$ on the STP and the SIR meta distribution ($\tau = 0 \, \mathrm{dB}$ and $x_0 = 0.9$).}
	\label{fig: Meta FigS_Opt_STPMeta_eta}
\end{figure}

Figs.~\ref{fig: Meta FigS_Opt_SingleMetric}-\ref{fig: Meta FigS_Opt_STPMeta_eta} provide some results obtained by solving Problem~\ref{problem: Meta general Prob} under both IN schemes. The default parameters are set as $x_0=0.9$, $\epsilon_\mathsf{R_c} = 5 \, \mathrm{m}$ for the fixed IN scheme (resp. $\epsilon_\mathsf{\mu} = 0.05 $ for the flexible IN scheme), $\epsilon_L = 1$, and $\epsilon_\xi = 0.05$. Fig.~\ref{fig: Meta FigS_Opt_SingleMetric} shows the results of performance metrics when the objective function $S$ in Problem~\ref{problem: Meta general Prob} is set as $p_s$, $1/V$, and $\bar{F}_{P_{s}}(x)$, respectively. From the figure, we can observe that in terms of maximizing the STP and the meta distribution, in the low SIR threshold region, the flexible IN scheme outperforms the fixed one, whereas in the high SIR threshold region, the fixed IN scheme performs better. When maximizing the metric $1/V$ (equivalently minimizing the variance), the flexible IN scheme is always a better choice (in the high SIR threshold region, the variance of the flexible scheme is slightly smaller than that of the fixed one).

Fig.~\ref{fig: Meta FigS_Opt_SingleMetric} discloses the information on which IN scheme is better when optimizing the three single metrics alone. Whereas Fig.~\ref{fig: Meta SingleOptimizationDesignPara} further presents the corresponding results of $( \mathsf{R_I} ^\star, L ^\star, \xi ^\star)$. From Fig.~\ref{fig: Meta SingleOptimizationDesignPara}\subref{fig: Meta sub FigS_Opt_SingleMaxqPara}, we can see that when maximizing the STP, in the low SIR threshold region (from Fig.~\ref{fig: Meta FigS_Opt_SingleMetric} we know that the flexible IN scheme performs better in this case), the flexible IN scheme tends to choose the largest $L$ and the largest $\xi$;\footnote{In our network setup, the ranges of the three parameters are respectively $\mu \in [0,5]$ (correspondingly $R_c \in [0, 446] \,\mathrm{m}$), $L \in [0,7]$, and $\xi \in [1,2.5]$. From simulations, we found that for the variance minimization problem, given a range of $\mu$, when $\tau$ is greater than a threshold (e.g., $5 \, \mathrm{dB}$ in Fig.~\ref{fig: Meta SingleOptimizationDesignPara}), the minima of $V$ is always reached at the upper end of the range. Therefore, to facilitate the solving of the problem, we set $\mu_{\max} = 5$ here. The relative margin gain of $V$ by adjusting $\mu_{\max} = 5$ to $\mu_{\max} = 6$ is no more than $2.5\%$.} whereas in the high SIR threshold region, the fixed IN scheme leads to the largest $L$ but smallest $\xi$. Fig.~\ref{fig: Meta SingleOptimizationDesignPara}\subref{fig: Meta sub FigS_Opt_SingleMinVarPara} shows that to minimize the variance, we need to choose the largest $L$ and $\xi$ for all SIR threshold region. From Fig.~\ref{fig: Meta SingleOptimizationDesignPara}\subref{fig: Meta sub FigS_Opt_SingleMaxMetaPara}, we see that to maximize the fraction of users with link reliability no less than $0.9$, in the low SIR threshold region, the flexible IN scheme leads to the largest $L$ and $\xi$; whereas in the high SIR threshold region, the fixed IN scheme leads to the largest $L$ but smallest $\xi$. Moreover, for each of the aforementioned single-metric optimization problems, an optimal $\mathsf{R_I}$ should be chosen to maximize the objective function.

Fig.~\ref{fig: Meta FigS_Opt_STPMeta_eta} shows the relationships between the weight coefficient $\eta$ and the STP as well as the SIR meta distribution under both IN schemes when considering the weighted metric optimization problem. The SIR threshold is set as $\tau = 0 \, \mathrm{dB}$. From the figure we observe that when setting the weighted metric $\eta p_s + (1 - \eta)  \bar{F}_{P_{s}}(x_0)$ as the objective function in Problem~\ref{problem: Meta general Prob}, the corresponding STP increases with $\eta$ while the corresponding SIR meta distribution decreases with $\eta$. This is because when $\eta$ increases, Problem~\ref{problem: Meta general Prob} focuses more on improving the STP, leading to a larger $p_s$ and a smaller $\bar{F}_{P_{s}}(x_0)$. Moreover, we can see that given the current values of $\tau$ and $x_0$, the flexible IN scheme outperforms the fixed one on both the STP and the meta distribution.

\begin{figure}[!t]	
	\centering
	\subfloat[Different Zipf exponents $\gamma_z$] {\includegraphics[scale=0.33]{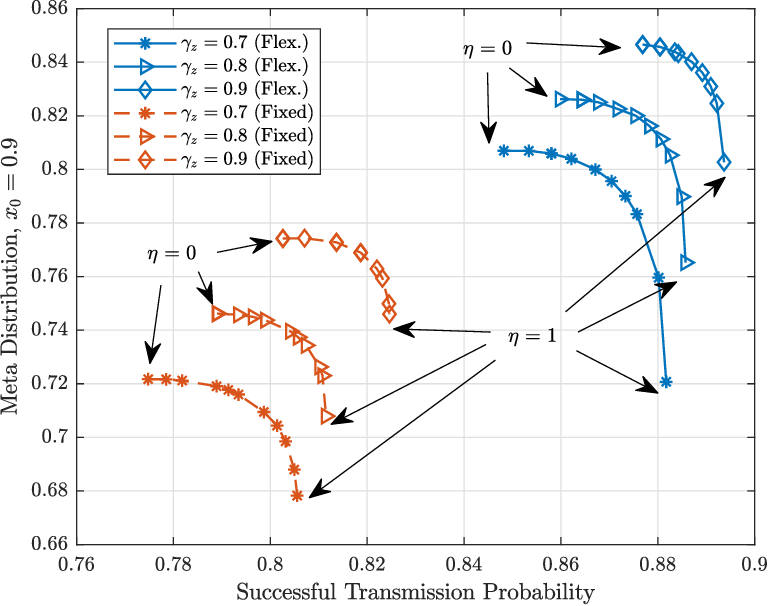} \label{fig: Meta sub FigS_Opt_Tradeoff_3Ga_eta} }
	\subfloat[Different cache sizes $C$] {\includegraphics[scale=0.33]{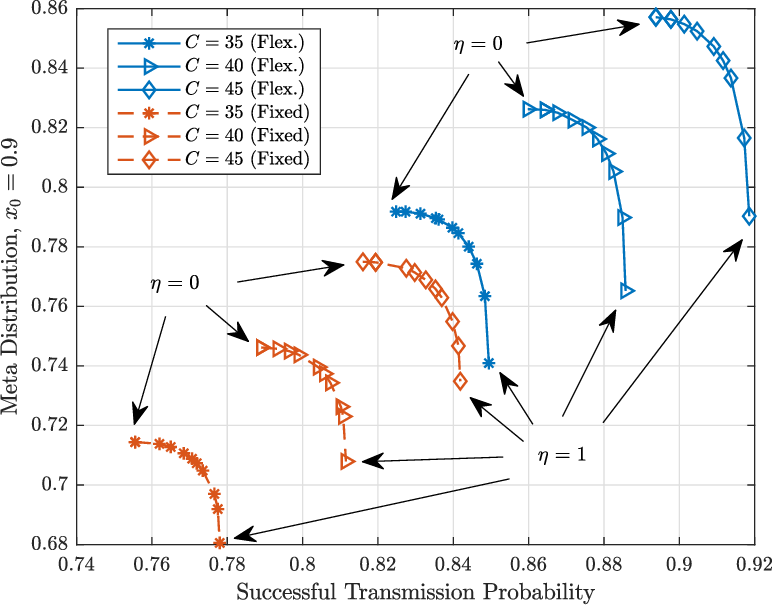} \label{fig: Meta sub FigS_Opt_Tradeoff_3C_eta}}
	\caption{Trade-off between the STP and the meta distribution with different Zipf exponents $\gamma_z$ and cache sizes $C$ ($\tau = 0 \, \mathrm{dB}$ and $x_0 = 0.9$).}
	\label{fig: Meta TradeoffBetweenSTPandMeta}
\end{figure}

Fig.~\ref{fig: Meta TradeoffBetweenSTPandMeta} depicts the trade-off between $p_s$ and $\bar{F}_{P_{s}}(x_0)$ with different Zipf exponents $\gamma_z$ and cache sizes $C$. The curves are obtained by tuning the weight coefficient $\eta$ from $0$ to $1$. The SIR threshold is set as $\tau = 0 \, \mathrm{dB}$. From Fig.~\ref{fig: Meta TradeoffBetweenSTPandMeta}\subref{fig: Meta sub FigS_Opt_Tradeoff_3Ga_eta}, we see that under both IN schemes, both the STP and the meta distribution increase when $\gamma_z$ increases, which means a more concentrated user preference for files leads to better network performance. From Fig.~\ref{fig: Meta TradeoffBetweenSTPandMeta}\subref{fig: Meta sub FigS_Opt_Tradeoff_3C_eta}, we see that under both IN schemes, increasing the cache size leads to higher $p_s$ and $\bar{F}_{P_{s}}(x_0)$, since more files can be stored in the network.


\subsection{System Design Guidelines}
Based on the analysis given in Section~\ref{subsec: Meta Numerical Effect of epsilon}, Section~\ref{subsec: Meta Numerical Effect of L}, and Section~\ref{subsec: Meta Numerical Effect of Xi} as well as the simulations in Section~\ref{subsec: Meta performance optimization}, we can obtain some observations for the following purposes:
\begin{enumerate}
	\item \textit{To improve the STP:} in the low SIR threshold region, it is better to choose the flexible IN scheme and cache more files in the network; whereas the fixed IN scheme with fewer files cached in the network performs better in the high SIR threshold region. For both cases, we should allocate more DoF for IN.
	\item \textit{To improve the link fairness:} the flexible IN scheme is always a better choice to improve the link fairness among different users, compared with the fixed one. Moreover, caching more files in the network and allocating more DoF for IN are more beneficial to the fairness improvement.
	\item \textit{To increase the fraction of users with high link reliability:} in the low SIR threshold region, we should choose the flexible IN scheme with more files cached in the network and larger DoF for IN; whereas in the high SIR threshold region, the fixed IN scheme with fewer files cached in the network and more DoF for IN is better. 
\end{enumerate}

Note that for all the purposes above, the optimal $\mathsf{R_I}$ can only be obtained by solving Problem~\ref{problem: Meta general Prob}. Furthermore, the numerical results obtained in Section~\ref{subsec: Meta performance optimization} further confirm the observations above.

\section{Conclusion}\label{section: Meta Conclusion}
In this paper, we investigate the SIR meta distribution for a multi-antenna cache-enabled network with two IN schemes considered, i.e., the fixed IN scheme and the flexible IN scheme. Using stochastic geometry analysis techniques, the expression for STP under each scheme is firstly derived. Then, we provide an approximated expression for the SIR meta distribution for each IN scheme, based on the first and second moments of a tight upper bound on the CSTP and utilizing the beta distribution. This is the first work to analyze the SIR meta distribution in a multi-antenna cache-enabled network incorporating the caching parameter and the IN parameters. It is shown that these two groups of parameters impact the system performance in a complicated manner, and the STP and the SIR meta distribution respond to the changes of parameters differently. Some useful system design insights are obtained by analysis and numerical simulation.

\appendices

\section{Derivation of $\bar{\varTheta}$}\label{appendix: Meta varThetabar}
\subsection{Flexible IN Scheme}
First, we consider the flexible IN scheme. The derivation procedure is similar to our previous work \cite[Appendix B]{Feng2022}. 
For completeness, we state the detailed procedures here. 

Let $\tilde{\Phi}_{u,n}$ and $\tilde{\lambda}_{u,n}$ respectively denote the set of served users that request file $n$ and the density of $\tilde{\Phi}_{u,n}$. Then $\tilde{\Phi}_{u,n}$ is an independent thinning of $\tilde{\Phi}_{u}$, since the file choices are independently made. Given the file diversity gain $\xi$, each BS randomly chooses $C$ different files from all the files in $\mathcal{N}_c$ with the same probability $T_c = 1/\xi$ to form a file combination. The set of all the file combinations is denoted by $\mathcal{I}$. $\mathcal{I}_i$ denotes the $i$-th combination from $\mathcal{I}$. Denote by $\mathcal{I}^n$ the set of file combinations containing file $n$. $ \mathcal{I}_i^n$ denotes the $i$-th combination from $\mathcal{I}^n$. Let $p_i$ be the probability that a BS stores combination $i$. Then, we have $T_c = \sum_{i \in \mathcal{I}^n } p_i $. The density of BSs that store file $n$ is $\lambda T_c $. Consider a typical BS which is located at the origin, denoted by $B_0$, and storing file $n$. Then, $B_0$ will store the combination $ \mathcal{I}_i^n$ with probability ${p_i \over T_c}$. Let $p(n \mid \mathcal{I}_i^n )$ denote the probability that the served user of a BS storing combination $ \mathcal{I}_i^n$ requests file $n$. To facilitate the analysis, we use the uniform distribution to approximate it, i.e., $p(n \mid \mathcal{I}_i^n ) \approx 1/C$, whose accuracy has been verified in \cite[Fig. 8]{Feng2022}. Then, we have $\tilde{\lambda}_{u,n} \approx \lambda T_c \sum_{i \in \mathcal{I}^n } {p_i \over T_c} p(n \mid \mathcal{I}_i^n ) = {\lambda T_c \over C} = {\lambda \over C \xi} $.

Next, we derive the expression for $\bar{\varTheta}^{\mathrm{fl}}$. We first characterize the probability that a served user sends an IN request to a BS. Consider a BS $B_0$ located at the origin and storing the file combination $\mathcal{I}_i$; and a user $u_x$ located at $x$ requesting file $n\in \mathcal{N}_c$. Denote by $Z$ the serving distance of the user $u_x$. 

Consider the case when $\mu <1$. When $0\leq \|x\| \leq \mu Z $, $u_x$ will send an IN request to $B_0$ if $n \notin \mathcal{I}_i$; otherwise, $B_0$ is the serving BS of $u_x$, which is contradictory. Denote by $ \tilde{p} (x, n \mid \mathcal{I}_i)$ the probability that $u_x$ sends an IN request to $B_0$. Then, we have $\tilde{p} (x, n \mid \mathcal{I}_i) = \mathds{1} \{ n \notin \mathcal{I}_i \}  \mathbb{P} \left[ \| x \| \leq \mu Z  \right]  =  \mathds{1} \{ n \notin \mathcal{I}_i \} \int_{\frac{\|x\|}{\mu}}^{\infty} f_{Z}(z) \mathrm{d}z	=  \mathds{1} \{ n \notin \mathcal{I}_i \} e^{-\pi \lambda \frac{\|x\|^2}{\xi \mu ^2}  } $, where $f_{Z} (z) = 2 \pi {\lambda \over \xi} z \exp (-\pi{\lambda \over \xi} z^2)$ is the PDF of the serving distance of $u_x$. Then, all the served users that request file $n$ and send an IN request to $B_0$ form an inhomogeneous PPP with density $\tilde{p} (x, n \mid \mathcal{I}_i) \tilde{\lambda}_{u,n}$. Therefore, the average number of IN requests sent by these users and received by $B_0$, denoted by $\bar{\varTheta}_{i,n}^{\mathrm{fl}}$, can be obtained by Campell's Theorem as $\bar{\varTheta} _ {i,n}^{\mathrm{fl}} =  \int_{\mathbb{R}^{2}} \tilde{p} (x, n \mid \mathcal{I}_i) \tilde{\lambda}_{u,n} \mathrm{d}x ={\lambda \over C \xi }  \int_{0}^{2\pi} \int_{0}^{\infty} \mathds{1} \{ n \notin \mathcal{I}_i \} e^{-\pi \lambda \frac{r^2}{\xi \mu ^2}  } r \mathrm{d}r \mathrm{d} \theta =  \mathds{1} \{ n \notin \mathcal{I}_i \} { \mu^2 \over C}$.
By summing $\bar{\varTheta} _ {i,n}^{\mathrm{fl}}$ for all the files in $\mathcal{N}_c$, we have $\bar{\varTheta} _ {i} ^{\mathrm{fl}}= \sum_{n \in \mathcal{N}_c} \bar{\varTheta} _ {i,n}^{\mathrm{fl}}
=  \left( \sum_{n \in \mathcal{N}_c} 1 - \sum_{n \in \mathcal{I}_i} 1 \right)  {\mu^2 \over C} $, which is the average number of IN requests received by $B_0$ on the condition that it stores file combination $\mathcal{I}_i$. By the total probability theorem, we have
\begin{align}\label{equ: Meta Appendix bar varTheta mu<1}
\bar{\varTheta}^{\mathrm{fl}}  & = \sum_{ i \in \mathcal{I}} p_i \bar{\varTheta} _ {i} ^{\mathrm{fl}}=
\sum_{ i \in \mathcal{I}} p_i \! \left( \sum_{n \in \mathcal{N}_c} 1 - \sum_{n \in \mathcal{I}_i} 1 \right)  { \mu^2 \over C} \notag
\\&= \left( N_c - \sum_{ i \in \mathcal{I}} \sum_{n \in \mathcal{I}_i} p_i \right) { \mu^2 \over C} \overset {(a)}{=}  \left(  N_c -  \sum_{ n \in \mathcal{N}_c}  T_c   \right)   { \mu^2 \over C} \notag
\\ & =  \left( N_c \! - \! C \right)   {\mu^2 \over C} = \xi\mu^2 - \mu^2 ,
\end{align}				
where (a) is from the relation $C = \sum_{i \in \mathcal{I}} p_i C = \sum_{i \in \mathcal{I}}  \sum_{ n \in \mathcal{I}_i} p_i \mathds{1} \{ n \in  \mathcal{I}_i  \}= \sum_{ n \in \mathcal{N}_c}\sum_{ i \in \mathcal{I}^n} p_i =\sum_{ n \in \mathcal{N}_c} T_c $ \cite{Cui2016}.

When $\mu\geq 1$, the BS $B_0$ will receive an IN request from $u_x$ if: 1) $0\leq \|x\| \leq Z $ and $n \notin \mathcal{I}_i$; 2) $Z < \|x\| \leq \mu Z $. Therefore, we have $\tilde{p} (x, n \mid \mathcal{I}_i) = \mathds{1} \{ n \notin \mathcal{I}_i \}  \mathbb{P} \left[ \| x \| \leq  Z  \right] +  \mathbb{P} \left[ Z < \| x \| \leq \mu Z  \right]=  \left( \mathds{1} \{ n \notin \mathcal{I}_i \} - 1\right) e^{ -\pi {\lambda \over \xi} \|x\|^2 } +  e^{-\pi \lambda \frac{\|x\|^2}{ \xi \mu ^2} }  $. Similarly, we have $\bar{\varTheta} _ {i,n}^{\mathrm{fl}}   = \int_{\mathbb{R}^{2}} \tilde{p} (x, n \mid \mathcal{I}_i) \tilde{\lambda}_{u,n} \mathrm{d}x = \left(  \mathds{1} \{ n \notin \mathcal{I}_i \} \! - \!1 \right) {1 \over C}  +  { \mu^2 \over C}  $, and 
\begin{equation}\label{equ: Meta Appendix bar varTheta mu>1}
\bar{\varTheta}^{\mathrm{fl}} =\sum_{ i \in \mathcal{I}} p_i \sum_{n \in \mathcal{N}_c} \bar{\varTheta} _ {i,n} ^{\mathrm{fl}}  =   \xi \mu^2 - 1.
\end{equation}
Combining \eqref{equ: Meta Appendix bar varTheta mu<1} and \eqref{equ: Meta Appendix bar varTheta mu>1}, we have the final result for $\bar{\varTheta}^{\mathrm{fl}} $.

\subsection{Fixed IN Scheme}
Here, we consider the fixed IN scheme. In this case, for an arbitrarily selected BS $B_0$, suppose the serving distance of its served user is $Z$. If $Z>R_c$, all the scheduled users within the circle of a radius $R_c$ centered at $B_0$ will send an IN request to $B_0$. Since each BS only serves one user at each time, $B_0$ will averagely receive $\pi  R_c^2\lambda$ IN requests. Nevertheless, if $Z \leq R_c$, $B_0$ will receive the IN requests from all the scheduled users within $R_c$ except its served user, and thus, the average number of IN requests is $\pi  R_c^2\lambda-1 $. Therefore, we have
\begin{align}\label{equ: Meta meanTheta Rc}
\bar{\varTheta}^\text{fx} &= \mathbb{P} [Z \leq R_c ] (\pi \lambda R_c^2 -1) + \mathbb{P} [Z > R_c ] \pi \lambda R_c^2 \notag
\\&= (\pi \lambda R_c^2 -1) \int_{0}^{R_c} f_Z(z) \mathrm{d}z +  \pi \lambda R_c^2 \int_{R_c}^{\infty} f_Z(z) \mathrm{d}z \notag
\\& =  \pi \lambda R_c^2 + e^{-\pi \frac{\lambda} {\xi} R_c^2} - 1,
\end{align}
where $f_{Z} (z) = 2 \pi {\lambda \over \xi} z \exp (-\pi{\lambda \over \xi} z^2)$ is the PDF of $Z$.

Finally, we obtain the results in \eqref{equ: Meta meanTheta} for the two schemes.

\section{Derivation of the IN Missing Probability}\label{appendix: Meta epsilon}
Denote by $\varepsilon^\prime (\varTheta)$ the probability that an arbitrary interfering BS $B_0$ within the IN range of $u_0$ does not select $u_0$ for IN when it has $\varTheta$ more IN requests besides the one from $u_0$. If $\varTheta+1\leq L$, $\varepsilon^\prime(\varTheta) = 0 $; if $\varTheta+1 > L$, $L$ IN requests are randomly uniformly chosen to be satisfied, and thus, $\varepsilon^\prime(\varTheta) = 1- {L \over \varTheta+1} $. Since whether or not a served user sends IN request to its interfering BSs is independent of others, given $B_0$ has received the IN request from $u_0$, $\varTheta$ follows the same PMF as in \eqref{equ: Meta PMF theta}. Therefore, we have $\varepsilon (\mathsf{R_I}, L,\xi) = \sum_{ \theta = L }^{\infty} \varepsilon ^\prime(\theta) \mathbb{P} \left[ \varTheta_r = \theta  \right]$. Further considering \eqref{equ: Meta PMF theta}, we have the final result.

\section{Derivation of the Expressions for the STPs}\label{appendix: Meta expression STP}
\subsection{Fixed IN Scheme}
We first consider the fixed IN scheme. The derivation is partly similar to \cite[Appendix C]{Feng2022}. The main differences come from that: 1) in \eqref{equ: Meta Appendix psn fx}, we need to consider the PMF $\mathbb{P} \left[ \varTheta_I^\text{fx} = \theta \right]$; 2) when calculating $\mathbb{P} [ g_{x_0} \geq \tau z^{\alpha} I  ] $ in \eqref{equ: Meta Appendix psn fx}, the densities $\lambda_{i}^{\text{fx}}$ and intervals $\varOmega_{i} ^{\text{fx} }$, $i\in\{a,b,c\}$ are different, as shown in \eqref{equ: Meta Appendix Lap_I prime}; and 3) the integration in \eqref{equ: Meta Appendix psn fx} cannot be written in a closed form, and is expressed as \eqref{equ: Meta psn fx}. We present the whole procedure for completeness. Based on the total probability formula, we have 
\begin{align}\label{equ: Meta Appendix psn fx}
p_{s,n}^\text{fx} \! &= \! \sum_{\theta = 0}^{ L  } \! \mathbb{P} \left[ \varTheta_I^\text{fx} = \theta \right] \!\!  \int_{0}^{\infty} \!\!  \mathbb{P} [ \Upsilon_{n}^\text{fx}  \geq \tau \!\! \mid \! \varTheta_I^\text{fx} = \theta, Z = z ] f_{Z} (z) \mathrm{d}z \notag
\\& = \! \sum_{\theta = 0}^{ L  } \! \mathbb{P} \left[ \varTheta_I^\text{fx} = \theta \right] \!\!  \int_{0}^{\infty} \!\!  \mathbb{P} [ g_{x_0} \geq \tau z^{\alpha} I  ] f_{Z} (z) \mathrm{d}z
,
\end{align}	
where  $\mathbb{P} [ \Upsilon_{n}^\text{fx}  \geq \tau \!\! \mid \! \varTheta_I^\text{fx} = \theta, Z = z ]$ is the STP conditioning on the number of IN requests satisfied by the serving BS of $u_0$ is $\varTheta_I^\text{fx} = \theta $, and the serving distance of $u_0$ is $Z=z$; $f_Z(z) = 2 \pi {\lambda \over \xi} z \exp (-\pi{\lambda \over \xi} z^2)$ is the PDF of $Z$; and $I \triangleq \sum_{x \in \Phi^\prime } g_{x} \|x\|^{-\alpha}$, with $\Phi ^\prime = \Phi_a^\text{fx} \cup \Phi_b^\text{fx} \cup \Phi_c^\text{fx}$, where $\Phi_a^{\text{fx}}$, $\Phi_b^{\text{fx}}$, and $\Phi_c^{\text{fx}}$ are given in \eqref{equ: Meta Phi a b c Rc}. Let $s \triangleq \tau z^{\alpha}  $, and $\mathcal{L}_{I}(s) \triangleq \mathbb{E}_I [ e ^{-sI}]$ denote the Laplace transform of $I$. Since $ g_{x_0} \overset{d}{\sim} \Gamma( M - \theta  , 1)$ is a gamma distributed variable, we have $\mathbb{P} \left[  g_{x_0} \geq \tau z^{\alpha} I  \right] =  \mathbb{E}_{I } \left[\sum_{m=0}^{ M - \theta -1} \frac{s^{m}}{m !}I^{ m} e^{-s I }\right] \overset{(a)}{=} \sum_{m=0}^{ M -  \theta -1} \frac{(-s)^{m}}{m !} \mathcal{L}_{I}^{(m)}(s)$ \cite{Li2015a}, where $\mathcal{L}_{I}^{(m)}(s)$ is the $m$-th derivative of $\mathcal{L}_{I}(s)$, and (a) is based on the property $\mathbb{E}_{I} \left[I^{ m} e^{-s I}\right]=(-1)^{m} \mathcal{L}_{ I} ^{(m)} (s)$. We further have
%
\begin{align}\label{equ: Meta Appendix Lap_I prime}
& \mathcal{L}_{I} \left( s \right)=  \mathbb{E}_{I} \left[ e^{-s I } \right]  = \mathbb{E}_{I} \!\left[ e^{ - s \sum_{x \in \Phi^\prime }  g_{x} \|x\|^{-\alpha }}   \right] \notag
\\& =  \mathbb{E}_{ \Phi^\prime } \left[\prod_{x \in \Phi^\prime } \mathbb{E}_{ g_{x}} \left[ e ^{ -s g_{x} \| x \| ^{-\alpha} } \right] \right] \notag \overset{(a)}{=}  \mathbb{E}_{ \Phi^\prime } \left[ \prod_{x \in \Phi^\prime} {1 \over  1 + s \| x \|^{-\alpha} }  \right] \notag
\\& \overset{(b)}{=} \exp \Bigg( \!\! \underbrace{ -2\pi \sum_{i = a}^c \lambda_{i}^{\text{fx}} \int_{\varOmega_{i} ^{\text{fx} } } \left( 1- {1 \over 1 + s v^{-\alpha} } \right) v \mathrm{d}v }_{\triangleq \chi (s)} \!\Bigg ),
\end{align}
where (a) is due to $ g_{x} \overset{d}{\sim}\text{Exp}(1)$, (b) is from the probability generating functional (PGFL) for a PPP and the polar-Cartesian coordinate transformation, and the densities $\lambda_{i}^{\text{fx}}$ and intervals $\varOmega_{i} ^{\text{fx} }$, $i\in\{a,b,c\}$ are given in \eqref{equ: Meta density of Phi a b c} and \eqref{equ: Meta Omega abc Rc}. The exponent $\chi(s)$ in \eqref{equ: Meta Appendix Lap_I prime} can be further obtained by
\begin{align}\label{equ: Meta Appendix chi}
\chi(s)  = &  -2\pi \Bigg[ \lambda_{a}^{\text{fx}} \left( \int_{0}^{\infty}  h(v) \mathrm{d}v  - \int_{\min \{Z, R_c\}}^{\infty}  h(v) \mathrm{d}v  \right)  \notag
\\ & + \lambda_{b}^{\text{fx}} \left( \int_{\min \{Z, R_c\}}^{\infty}  h(v) \mathrm{d}v  - \int_{\max \{Z, R_c \}}^{\infty}  h(v) \mathrm{d}v   \right) \notag
\\ & + \lambda_{c}^{\text{fx}} \int_{\max \{Z, R_c \}}^{\infty}  h(v) \mathrm{d}v  \Bigg],
\end{align}
where $h(v) \triangleq \left( 1- (1 + s v^{-\alpha_2})^{-1} \right) v$. Using \cite[(3.194)]{Zwillinger2014} and after some manipulation, we can obtain the expression for $q_0 = \chi(s) $ as shown in \eqref{equ: Meta q_0 fx}. Since $\mathcal{L}_{I} ^ \prime (s)= \mathcal{L}_{I} (s) \chi ^ \prime (s)$, according to the Leibniz formula, we have
\begin{equation}\label{equ: Meta Appendix Leibniz}
\mathcal{L}_{I}^{(m)}(s)=\sum_{k=0}^{m-1}\binom{m-1}{k} \chi^{(m-k)}(s) \mathcal{L}_{I}^{(k)}(s),
\end{equation}	
where $\chi^{(k)}(s)$ denotes the $k$-th derivative of $\chi(s)$ with respect to $s$ and is given by
\begin{align}\label{equ: Meta Appendix chi_k}
\chi ^{(k)}(s)  =    2\pi \sum_{i = a}^c \lambda_{i}^{\text{fx}} \int_{\varOmega_{i} ^{\text{fx} } } (-1)^k k! \frac{\left( v^{-\alpha} \right)^k v  }{ \left( 1 + s v^{-\alpha} \right)^{1 + k} }\mathrm{d}v.
\end{align}			
Let $q_m = \frac{(-s)^{m}}{(m) !} \chi^{(m)}(s)$, substituting $s = \tau z^\alpha$ into \eqref{equ: Meta Appendix chi_k}, considering \cite[(3.194)]{Zwillinger2014}, and after some manipulation, we can obtain the expression for $q_m$ given in \eqref{equ: Meta q_m fx}.

Let $x_{m}=\frac{1}{m !}(-s)^{m} \mathcal{L}_{I}^{(m)}(s)$. From \eqref{equ: Meta Appendix Leibniz}, we have $x_0 = \mathcal{L}_{I}(s) = \exp(q_0)$ and
\begin{align} \label{equ: Meta Appendix xm}
x_{m} & =\sum_{k=0}^{m-1} \frac{m-k}{m}\left(\frac{(-s)^{m-k}}{(m-k) !} \chi^{(m-k)}(s)\right) x_{k} \notag
\\& = \sum_{k=0}^{m-1} \frac{m-k}{m} q_{m-k} x_{k}.
\end{align}
To obtain $\mathbb{P} \left[  g_{x_0} \geq \tau z^{\alpha} I  \right] = \sum_{m=0}^{ M -  \theta -1} \frac{(-s)^{m}}{m !} \mathcal{L}_{I}^{(m)}(s) =  \sum_{m=0}^{ M -  \theta -1} x_m $, we need to solve the explicit expression for $x_m$. We define two power series as
\begin{equation}\label{equ: Meta Appendix Define of Q X}
Q(t) \triangleq \sum_{m=0}^{\infty} q_{m}  t^{m}, \quad X(t) \triangleq \sum_{m=0}^{\infty} x_{m} t^{m}.
\end{equation}
Consider that the derivative of a series $ Y(t) \triangleq \sum_{m=0}^{\infty} y_{m}  t^{m}$ is $Y^{(1)}(t)=\sum_{m=0}^{\infty} m y_{m}  t^{m-1}$, and the product of $Q(t)$ and $X(t)$ is $Q(t) X(t)=\sum_{m=0}^{\infty} (\sum_{i=0}^{m}  q_{m-i}  x_{i} ) t^{m}$. From \eqref{equ: Meta Appendix xm}, we can obtain
\begin{equation}\label{equ: Meta Appendix diff equ}
X^{(1)}(t) = Q^{(1)}(t) X(t).
\end{equation}
The solution of the above differential equation is $X(t)=c \exp \left( Q(t)\right)$. Since $X(0) = x_0 = \mathcal{L}_{I} (s)$ and $Q(0) = q_0 $, we have $c = 1$. Thus, the solution of \eqref{equ: Meta Appendix diff equ} is $X(t)= \exp \left( Q(t)\right)$. Since $\mathbb{P} \left[  g_{x_0} \geq \tau z^{\alpha} I  \right]  =  \sum_{m=0}^{ M -  \theta -1} x_m $, we have
\begin{align}\label{equ: Meta Appendix Coefficients}
\mathbb{P} \left[  g_{x_0} \geq \tau z^{\alpha} I \right] & =  \sum_{m=0}^{ M -  \theta -1} x_m = \sum_{m=0}^{ M -  \theta -1} \left. \frac{1}{m!} X^{(m)}(t)\right|_{t=0} \notag
\\& = \sum_{m=0}^{ M -  \theta -1} \left.\frac{1}{m !} \frac{\mathrm{d}^{m}}{\mathrm{d} t^{m}} \exp \left( Q(t)\right) \right|_{t=0}.
\end{align}
From \cite[p.~14]{henrici1993applied} and \cite[Appendix A]{Li2014}, it can be shown that the first $ M -  \theta $ coefficients of the power series $\exp \left( Q(z)\right)$ is the first column of the matrix exponential $\exp \left( \mathbf{Q}_{M -  \theta } (R_c, L, \xi) \right) $, and their sum equals the $L_1$ induced norm of this matrix, where $\mathbf{Q}_{D} (R_c, L, \xi)$ is given in \eqref{equ: Meta matrixQ fx}. Further considering \eqref{equ: Meta Appendix psn fx}, we obtain the final result as in \eqref{equ: Meta psn fx}.
\subsection{Flexible IN Scheme}
For the flexible IN scheme, the derivation is similar to the fixed IN scheme. However, in \eqref{equ: Meta Appendix psn fx}, we need to consider the PMF $\mathbb{P} \left[ \varTheta_I^\text{fl} = \theta \right]$; in \eqref{equ: Meta Appendix Lap_I prime} and \eqref{equ: Meta Appendix chi_k}, the integration intervals $\varOmega_{i} ^{\text{fl} }$ and densities $\lambda_{i}^{\text{fl}}$, $i\in\{a,b,c\}$ should be considered. In this case, we substitute the series $Q(t)$ in \eqref{equ: Meta Appendix Define of Q X} with $\tilde{ W } (t) \triangleq \sum_{m=0}^{\infty} w_{m}  t^{m}$, and the corresponding coefficients $w_m$, $m=0,1,2,\dots$, are given by \eqref{equ: Meta w0 fl} and \eqref{equ: Meta wm fl}. Then, the corresponding solution to \eqref{equ: Meta Appendix diff equ} becomes $X(t) = \exp ( \pi z^{2} \lambda (   \tilde{ W }(t) -2w_0 ) )$, with which the integration in \eqref{equ: Meta Appendix psn fx} can be written in a closed form as
\begin{align}\label{equ: Meta Appendix Psi2Theta_FinalExpression fl}
&\int_{0}^{\infty} \!\!  \mathbb{P} [  g_{x_0} \geq \tau z^{\alpha} I  ] f_{Z} (z) \mathrm{d}z \notag
=  \int_{0}^{\infty}  \sum_{m=0}^{ M -  \theta -1} x_m f_{Z} (z) \mathrm{d}z \notag
\\ = & \int_{0}^{\infty} \sum_{m=0}^{M -  \theta -1} \left. \frac{1}{m!} \frac{\mathrm{d}^{m}}{\mathrm{d} t^{m}} X(t)\right|_{t=0} f_{Z} (z) \mathrm{d}z \notag
\\ = &  \sum_{m=0}^{M -  \theta -1} \left. \frac{1}{m!} \frac{\mathrm{d}^{m}}{\mathrm{d} t^{m}} \! \int_{0}^{\infty} \!\! \! \! \exp \! \left( \pi z ^{2} \lambda \left(   \tilde{ W }(t) \!- \! 2w_0 \right) \! \right) \! f_{Z } (z ) \mathrm{d}z \right|_{t=0} \notag
\\ = & { 1 \over \xi  } \sum_{m=0}^{M -  \theta -1} \left.\frac{1}{m !} \frac{\mathrm{d}^{m}}{\mathrm{d} t^{m}} \left( { 1 \over \xi  } + 2 w_0  - \tilde{ W } (t) \right) ^{-1}  \right|_{t=0} \notag
\\ = & { 1 \over \xi  } \left\| \left[  \left( { 1 \over \xi  } + 2 w_0  \right) \mathbf{I}_{M -  \theta} - \tilde{ \mathbf{W} } _{M -  \theta}(\mu, L, \xi) \right] ^{-1} \right\|_1,
\end{align}
which finishes the derivation.

\section{Derivation of the Upper Bound $ P_{s,n}^u(\tau)$}\label{appendix: Meta Upper Bound}
Based on the total probability formula, we have $P_{s,n}(\tau)  = \mathbb{P}\left[ \Upsilon_n  \geq \tau \mid \Phi \right] = \sum_{\theta = 0}^{L} \mathbb{P} \left[ \varTheta_I = \theta \right] \mathbb{P} \left[  \Upsilon_n \geq \tau \mid \Phi, \varTheta_I = \theta  \right]$,
where $ \mathbb{P} \left[  \Upsilon_n \geq \tau \mid \Phi, \varTheta_I = \theta  \right] $ is the CSTP with further condition that the number of IN requests satisfied by the serving BS of $u_0$ is $\varTheta_I = \theta$. Then, it remains to calculate $ \mathbb{P} \left[  \Upsilon_n \geq \tau \mid \Phi, \varTheta_I = \theta  \right] $. Let $I = \sum_{x \in \Phi^\prime  }    g_{x} \|x\|^{-\alpha }$ be the interference received at $u_0$, where $\Phi^\prime \triangleq \Phi_a  \cup \Phi_b  \cup \Phi_c $ denotes all the interfering BSs of $u_0$. According to \eqref{equ: Meta SIR expression}, we have
	\begin{align}\label{equ: Meta Appendix Upper Bound}
&\mathbb{P} \left[  \Upsilon_n \geq \tau \mid \Phi, \varTheta_I = \theta \right] \notag
\\ & =  \mathbb{E}_{g_x} \left[ \mathbb{P} \left[  g_{x_0} \geq \tau Z^{\alpha} I \mid \Phi, \varTheta_I = \theta \right] \right] \notag
\\&\stackrel{(a)}{\leq} 1- \mathbb{E}_{g_x} \left[ \left( 1- \exp \left(-\beta \tau Z^{\alpha} I\right) \right)^{M - \theta } \right] \notag
\\& = 1- \mathbb{E}_{g_x} \left[ \sum_{i=0}^{M-\theta } (-1)^{i} \binom{M-\theta }{i} \exp \left(- i \beta \tau Z^{\alpha} I\right)   \right] \notag
\\& = \sum_{i=1}^{M-\theta } (-1)^{i+1} \binom{M-\theta }{i} \mathbb{E}_{g_x}\! \!\left[\!\exp\! \left( \!\!- i \beta \tau Z^{\alpha} \sum_{x \in \Phi^\prime }  g_{x} \|x\|^{-\alpha}\!\! \right) \!\!\right] \notag
\\& \overset{(b)}{=} \sum_{i=1}^{ M-\theta } (-1)^{i+1} \binom{ M-\theta }{i} \prod_{x \in \Phi^\prime} \mathbb{E}_{g_x} \!\! \left[ \exp \left(- i \beta \tau Z^{\alpha}  g_{x} \|x\|^{-\alpha} \right) \right] \notag
\\& \overset{(c)}{=} \sum_{i=1}^{ M-\theta  } (-1)^{i+1} \binom{ M-\theta }{i} \prod_{x \in \Phi^\prime } {1 \over 1 + i \beta \tau Z^{\alpha} \|x\|^{-\alpha} }\notag
\\& \triangleq P_{s,n,\varTheta_I}^u(\tau, \theta),
\end{align}		
where $Z$ is the serving distance of $u_0$, given a realization of $\Phi$; (a) is due to the Gamma distributed wireless channel $g_{x_0} \overset{d}{\sim} \Gamma(M-\theta , 1)$ (whose CCDF is $\mathbb{P} [g_{x_0} > y] = 1- {\gamma (M-\theta,y ) \over \Gamma(M-\theta)}$), and a lower bound on the incomplete Gamma function ${\gamma (M-\theta,y ) \over \Gamma(M-\theta)}$, i.e., $\left( 1- e^{-\beta y} \right) ^ {M-\theta } \leq {\gamma( M-\theta,y) \over \Gamma( M - \theta )}$, with $\beta = ((M-\theta) !)^{ -1 / (M-\theta) }$; (b) is due to the independence of Rayleigh fading channels for different BSs; (c) is obtained by taking the expectation over the exponential distributed variable $g_x \overset{d}{\sim} \text{Exp}(1) $. This completes the derivation.

\section{Derivation of $M_{1,n}^u$ and $M_{2,n}^u$}\label{appendix: Meta M1nu M2nu}
\subsection{Fixed IN Scheme}
We first consider the flexible IN scheme. Based on $P_{s,n}^u(\tau)$ given in \eqref{equ: Meta Psn upper} and the definition of $P_{s,n,\varTheta_I}^u(\tau, \theta)$ in \eqref{equ: Meta Appendix Upper Bound}, we calculate the expression for $M_{1,n}^{\text{fx},u}$ as follows.
	\begin{align}\label{equ: Meta Appendix M1nu Rc}
& M_{1,n}^{\text{fx},u}   =  \mathbb{E} \left[ \sum_{\theta = 0}^{L} \mathbb{P} \left[ \varTheta_I ^{\text{fx}} = \theta \right]  P_{s,n,\varTheta_I^{\text{fx}} }^{\text{fx},u}(\tau, \theta) \right] \notag
\\ & = 	 \sum_{\theta = 0}^{L} \mathbb{P} \left[ \varTheta_I^{\text{fx}} = \theta \right]  \sum_{i=1}^{ M-\theta  } (-1)^{i+1} \binom{ M-\theta }{i} \notag
\\  &\qquad \qquad \quad \times \mathbb{E} \Bigg[ \prod_{x \in \Phi^\prime } {1 \over 1 + i \beta \tau Z^{\alpha} \|x\|^{-\alpha} } \Bigg] \notag
\\ & \overset{(a)}{ = } \sum_{\theta = 0}^{L} \mathbb{P} \left[ \varTheta_I^{\text{fx}} = \theta \right]  \sum_{i=1}^{ M-\theta  } (-1)^{i+1} \binom{ M-\theta }{i} \int_{0}^{\infty} f_{Z}(z) \notag
\\  & \quad  \times  \exp \Bigg( \!\!  -2\pi \sum_{k = a}^c \lambda_k^{\text{fx}} \int_{\varOmega_k^{\text{fx}}} \left( 1-    { 1 \over  1 + i \beta \tau z^{\alpha} v^{-\alpha}  }  \right) v  \mathrm{d}v
\!\Bigg ) \mathrm{d}z \notag 
\\ & \overset{(b)}{=} \sum_{\theta = 0}^{L} \mathbb{P} \left[ \varTheta_I^\text{fx} = \theta \right]  \sum_{i=1}^{ M-\theta  } (-1)^{i+1} \binom{ M-\theta }{i} \int_{0}^{\infty} f_{Z}(z) \notag
\\  &  \quad  \times  \exp \Bigg(   - \pi \lambda   \Bigg( \varepsilon^\text{fx} (R_c, L, \xi) \left(1- {1\over \xi } \right)  z^{2}   \frac{ ( i \beta \tau ) ^{ \frac{2} {\alpha}} }{\mathrm{sinc}(\frac{2} {\alpha })}   \notag
\\ & \quad  + a^\text{fx}{ z^2 \over \xi }  F ( i \beta \tau )  + b^\text{fx} \left(1- \varepsilon^\text{fx} (R_c, L, \xi) \right) R_c^2 \notag
\\ & \quad  \times F\left( i \beta \tau \left( { z \over R_c} \right)^{ \alpha} \right)     \Bigg)	\!\Bigg ) \mathrm{d}z,
\end{align}
where $\Phi^\prime = \Phi_a^{\text{fx}} \cup \Phi_b^{\text{fx}} \cup \Phi_c^{\text{fx}}$; (a) is due to the PGFL of a PPP \cite{Haenggi2012}; $f_{Z} (z) = 2 \pi {\lambda \over \xi} z \exp (-\pi{\lambda \over \xi} z^2)$ is the PDF of the serving distance $Z$ of $u_0$; $\lambda_k^{\text{fx}}$, $k \in \{ a,b,c \}$ are given in \eqref{equ: Meta density of Phi a b c}, and the integration limits $\varOmega_k^{\text{fx}}$, $k \in \{a,b,c\}$, are given in \eqref{equ: Meta Omega abc Rc}; (b) is from \cite[(3.194)]{Zwillinger2014}, and  $\mathrm{sinc}(x) = \frac{\sin \pi x}{\pi x}$.

For the second moment $M_{2,n}^{\text{fx},u}$, from \eqref{equ: Meta Psn upper} and the definition of $P_{s,n,\varTheta_I}^u(\tau, \theta)$ in \eqref{equ: Meta Appendix Upper Bound}, we have
\begin{align}\label{equ: Meta Appendix M2un}
& M_{2,n}^{\text{fx},u}=  \mathbb{E} \left[\left( \sum_{\theta = 0}^{L} \mathbb{P} \left[ \varTheta_I ^{\text{fx}}  = \theta \right]  P_{s,n,\varTheta_I^{\text{fx}} }^{\text{fx},u}(\tau, \theta) \right)^2 \right] \notag
\\ & = \int_{0}^{\infty} \!\! \mathbb{E}_{\Phi}  \!\! \left[\left( \sum_{\theta = 0}^{L} \mathbb{P} \left[ \varTheta_I ^{\text{fx}} = \theta \right]  P_{s,n,\varTheta_I^{\text{fx}} ,Z}^{\text{fx},u}(\tau, \theta, z) \right)^2 \right]   f_{Z}(z) \mathrm{d}z,
\end{align}
where $P_{s,n,\varTheta_I^{\text{fx}} ,Z}^{\text{fx},u}(\tau, \theta, z) \triangleq \mathbb{P} \left[  \Upsilon_n ^{\text{fx}} \geq \tau \mid \Phi, \varTheta_I^{\text{fx}}  = \theta, Z = z \right]$ is the upper bound of the CSTP with further conditions that $\varTheta_I ^{\text{fx}} = \theta$ and the serving distance of $u_0$ is $Z=z$. We then have
	\begin{align}\label{equ: Meta Appendix E Theta Psu}
&\mathbb{E}_{\Phi} \left[\left( \sum_{\theta = 0}^{L} \mathbb{P} \left[ \varTheta_I ^{\text{fx}} = \theta \right]  P_{s,n,\varTheta_I^{\text{fx}} ,Z}^{\text{fx},u}(\tau, \theta, z) \right)^2 \right]  \notag
\\& = \mathbb{E}_{\Phi} \Bigg[ \left( \sum_{\theta_1 = 0}^{L} \mathbb{P} \left[ \varTheta_I ^{\text{fx}} = \theta_1 \right]  P_{s,n,\varTheta_I^{\text{fx}} ,Z}^{\text{fx},u}(\tau, \theta_1, z) \right) \notag
\\& \qquad \qquad \times \left( \sum_{\theta_2 = 0}^{L} \mathbb{P} \left[ \varTheta_I ^{\text{fx}} = \theta_2 \right]  P_{s,n,\varTheta_I^{\text{fx}} ,Z}^{\text{fx},u}(\tau, \theta_2, z) \right) \Bigg] \notag
\\& = \mathbb{E}_{\Phi}\Bigg [   \sum_{\theta_1 = 0}^{ L }  \sum_{\theta_2 = 0}^{ L }   \mathbb{P} \left[ \varTheta_I ^{\text{fx}} = \theta_1 \right]  \mathbb{P} \left[ \varTheta_I^{\text{fx}}  = \theta_2 \right] \notag
\\& \qquad \qquad \qquad \quad \times  P_{s,n,\varTheta_I^{\text{fx}} ,Z}^{\text{fx},u}(\tau, \theta_1, z)  P_{s,n,\varTheta_I^{\text{fx}} ,Z}^{\text{fx},u}(\tau, \theta_2, z) \Bigg] \notag
\\& = \sum_{\theta_1 = 0}^{ L }  \sum_{\theta_2 = 0}^{ L } \mathbb{P} \left[ \varTheta_I^{\text{fx}}  = \theta_1 \right]  \mathbb{P} \left[ \varTheta_I ^{\text{fx}} = \theta_2 \right]  \notag
\\& \qquad   \quad \times \mathbb{E}_{\Phi}\left[  P_{s,n,\varTheta_I^{\text{fx}} ,Z}^{\text{fx},u}(\tau, \theta_1, z)  P_{s,n,\varTheta_I^{\text{fx}} ,Z}^{\text{fx},u}(\tau, \theta_2, z)\right],
\end{align}
where 
\begin{align}\label{equ: Meta Appendix E Psu}
& \mathbb{E}_{ \Phi }\left[  P_{s,n,\varTheta_I^{\text{fx}} ,Z}^{\text{fx},u}(\tau, \theta_1, z)  P_{s,n,\varTheta_I^{\text{fx}} ,Z}^{\text{fx},u}(\tau, \theta_2, z) \right] \notag
\\& = \mathbb{E}_{ \Phi }\Bigg[  \Bigg( \sum_{i=1}^{M-\theta_1} \! (-1)^{i+1} \binom{M-\theta_1}{i} \prod_{x \in \Phi^\prime } {1 \over 1 + i \beta_1 \tau z^{\alpha} \|x\|^{-\alpha} } \Bigg) \notag
\\& \quad \times \Bigg( \sum_{j=1}^{M - \theta_2 } (-1)^{j+1} \binom{M - \theta_2 }{j} \prod_{x \in \Phi^\prime } {1 \over 1 + j \beta_2 \tau z^{\alpha} \|x\|^{-\alpha} } \Bigg) \Bigg] \notag
\\& = \sum_{i=1}^{M-\theta_1} \sum_{j=1}^{M - \theta_2 } (-1)^{i+j} \binom{M-\theta_1}{i} \binom{M - \theta_2 }{j}  \notag
\\& \quad \times \mathbb{E}_{\Phi }\left[   \prod_{x \in \Phi^\prime} {1 \over ( 1 + i \beta_1 \tau z^{\alpha} \|x\|^{-\alpha} ) ( 1 + j \beta_2 \tau z^{\alpha} \|x\|^{-\alpha} ) }  \right] \notag
\\& \overset{(a)}{=} \sum_{i=1}^{M-\theta_1} \sum_{j=1}^{M - \theta_2 } (-1)^{i+j} \binom{M-\theta_1}{i} \binom{M - \theta_2 }{j} \exp \Bigg(   -2\pi  \notag
\\& \quad \times \sum_{k = a}^c \! \lambda_k^{\text{fx}} \!\! \int_{\varOmega_k^{\text{fx}}} \!\! \left( \! 1 \! - \!{ 1 \over ( 1 \!+ \! i \beta_1 \tau z^{\alpha} v^{-\alpha} ) ( 1 \! + \! j \beta_2 \tau z^{\alpha} v^{-\alpha} )  }  \right) \! v \mathrm{d}v \! \Bigg ) \notag
\\& = \sum_{i=1}^{M-\theta_1} \sum_{j=1}^{M - \theta_2 } (-1)^{i+j} \binom{M-\theta_1}{i} \binom{M - \theta_2 }{j} \notag
\\& \quad \times \exp \left(   -2\pi \sum_{k = a}^c \lambda_k^{\text{fx}}  \mathcal{H}_{ij} (\varOmega_k^{\text{fx}}, \beta_1 \tau z^{\alpha} , \beta_2 \tau z^{\alpha} ) \right ),
\end{align}
where $\beta_m= ( (M-\theta_m) !)^{-1\over M-\theta_m }$, $m \in \{1,2\}$, and (a) is from the PGFL of a PPP; $\mathcal{H}_{ij} (\varOmega, x , y )$ is given in \eqref{equ: Meta Hij}. Considering \eqref{equ: Meta Appendix M2un}, \eqref{equ: Meta Appendix E Theta Psu}, and \eqref{equ: Meta Appendix E Psu}, after some manipulation, we have the expression for $ M_{2,n}^{\text{fx},u}$ shown in \eqref{equ: Meta M2nu fx}.

\subsection{Flexible IN Scheme}
The derivation for the flexible IN scheme is similar to the fixed IN scheme. However, in \eqref{equ: Meta Appendix M1nu Rc}, we need to consider the PMF $\mathbb{P} \left[ \varTheta_I^\text{fl} = \theta \right]$, and the integration intervals $\varOmega_{i} ^{\text{fl} }$ and densities $\lambda_{i}^{\text{fl}}$, $i\in\{a,b,c\}$ should be considered. Correspondingly, the integration of (b) in \eqref{equ: Meta Appendix M1nu Rc} can be written as
	\begin{align}
&\sum_{\theta = 0}^{L} \mathbb{P} \left[ \varTheta_I^{\text{fl}} = \theta \right]  \sum_{i=1}^{ M-\theta  } (-1)^{i+1} \binom{ M-\theta }{i} \int_{0}^{\infty} f_{Z}(z) \notag
\\  & \quad  \times  \exp \Bigg(   -\pi z^2 \lambda \Big( \varepsilon^{\text{fl}} (\mu, L, \xi)  (1  - {1\over \xi } )   \frac{ ( i \beta \tau ) ^{ \frac{2} {\alpha}} }{\mathrm{sinc}(\frac{2} {\alpha })}   \notag
\\ & \quad +   a^{\text{fl}}  \left( 1  -   \varepsilon^{\text{fl}} (\mu, L, \xi) \right) \mu^2   F   (  { i \beta \tau \over \mu ^{ \alpha }}   )  +  b^{\text{fl}}  {1\over \xi } F( i \beta \tau )  \Big)
\!\Bigg ) \mathrm{d}z  \notag
\\& = \sum_{\theta = 0}^{L} \mathbb{P} \left[ \varTheta_I^{\text{fl}} = \theta \right]  \sum_{i=1}^{ M-\theta  } (-1)^{i+1} \binom{ M-\theta }{i}   \notag
\\& \quad \times \Big( 1+ \xi \Big(  \varepsilon^{\text{fl}} (\mu, L, \xi)  (1  - {1\over \xi } )   \frac{ ( i \beta \tau ) ^{ \frac{2} {\alpha}} }{\mathrm{sinc}(\frac{2} {\alpha })}   \notag
\\ & \quad +   a^{\text{fl}}  \left( 1  -   \varepsilon^{\text{fl}} (\mu, L, \xi) \right) \mu^2   F   (  { i \beta \tau \over \mu ^{ \alpha }}   )  +  b^{\text{fl}}  {1\over \xi } F( i \beta \tau )  \Big)
\!\Big ) ^{-1},
\end{align}
where the equality follows $\int_{0}^{\infty} 2xe^{-Ax^2} \mathrm{d}x = 1/A $. Moreover, following the same procedure as \eqref{equ: Meta Appendix M2un}-\eqref{equ: Meta Appendix E Psu}, we can obtain the expression for $M_{2,n}^{\text{fl},u}$.

%




\ifCLASSOPTIONcaptionsoff
\newpage
\fi
\bibliographystyle{IEEEtran}
\bibliography{IEEEabrv,mylib_abrv}

\begin{thebibliography}{10}
\providecommand{\url}[1]{#1}
\csname url@samestyle\endcsname
\providecommand{\newblock}{\relax}
\providecommand{\bibinfo}[2]{#2}
\providecommand{\BIBentrySTDinterwordspacing}{\spaceskip=0pt\relax}
\providecommand{\BIBentryALTinterwordstretchfactor}{4}
\providecommand{\BIBentryALTinterwordspacing}{\spaceskip=\fontdimen2\font plus
\BIBentryALTinterwordstretchfactor\fontdimen3\font minus
  \fontdimen4\font\relax}
\providecommand{\BIBforeignlanguage}[2]{{%
\expandafter\ifx\csname l@#1\endcsname\relax
\typeout{** WARNING: IEEEtran.bst: No hyphenation pattern has been}%
\typeout{** loaded for the language `#1'. Using the pattern for}%
\typeout{** the default language instead.}%
\else
\language=\csname l@#1\endcsname
\fi
#2}}
\providecommand{\BIBdecl}{\relax}
\BIBdecl

\bibitem{Dimakis2013}
K.~Shanmugam, N.~Golrezaei, A.~G. Dimakis, A.~F. Molisch, and G.~Caire,
  ``{FemtoCaching}: {W}ireless content delivery through distributed caching
  helpers,'' \emph{IEEE Trans. Inf. Theory}, vol.~59, no.~12, pp. 8402--8413,
  Dec. 2013.

\bibitem{Li2017a}
Q.~Li, W.~Shi, X.~Ge, and Z.~Niu, ``{C}ooperative edge caching in
  software-defined hyper-cellular networks,'' \emph{IEEE J. Sel. Areas
  Commun.}, vol.~35, no.~11, pp. 2596--2605, Nov. 2017.

\bibitem{Liu2016a}
D.~Liu, B.~Chen, C.~Yang, and A.~F. Molisch, ``{Caching at the wireless edge:
  Design aspects, challenges, and future directions},'' \emph{IEEE Commun.
  Mag.}, vol.~54, no.~9, pp. 22--28, Sep. 2016.

\bibitem{Kuang2019}
S.~Kuang, X.~Liu, and N.~Liu, ``{A}nalysis and optimization of random caching
  in \${K}\$-tier multi-antenna multi-user {HetNets},'' \emph{IEEE Trans.
  Commun.}, vol.~67, no.~8, pp. 5721--5735, Aug. 2019.

\bibitem{Kuang2019a}
S.~Kuang and N.~Liu, ``{A}nalysis and optimization of random caching in
  multi-antenna small-cell networks with limited backhaul,'' \emph{IEEE Trans.
  Veh. Technol.}, vol.~68, no.~8, pp. 7789--7803, Aug. 2019.

\bibitem{Jiang2019c}
D.~Jiang and Y.~Cui, ``{E}nhancing performance of random caching in large-scale
  wireless networks with multiple receive antennas,'' \emph{IEEE Trans.
  Wireless Commun.}, vol.~18, no.~4, pp. 2051--2065, Apr. 2019.

\bibitem{Xu2019c}
X.~Xu and M.~Tao, ``{M}odeling, analysis, and optimization of caching in
  multi-antenna small-cell networks,'' \emph{IEEE Trans. Wireless Commun.},
  vol.~18, no.~11, pp. 5454--5469, Nov. 2019.

\bibitem{Zhi2019}
K.~Zhi, G.~Chen, L.~Qiu, X.~Liang, and C.~Ren, ``{A}nalysis and optimization of
  random cache in multi-antenna {HetNets} with interference nulling,'' in
  \emph{2019 IEEE Glob. Commun. Conf.}, Dec. 2019, pp. 1--6.

\bibitem{Liu2021}
W.~Liu, L.~Li, L.~Jiao, H.~Dai, and G.~Zheng, ``{J}oint interference alignment
  and probabilistic caching in {MIMO} small-cell networks,'' \emph{IEEE Trans.
  Veh. Technol.}, vol.~70, no.~9, pp. 9400--9407, Sep. 2021.

\bibitem{Feng2022}
\BIBentryALTinterwordspacing
T.~Feng, X.~Gu, and B.~Liang, ``{R}andom caching design for multi-user
  multi-antenna {HetNets} with interference nulling,'' \emph{arXiv:2201.11879},
  Jan. 2022. [Online]. Available: \url{http://arxiv.org/abs/2201.11879}
\BIBentrySTDinterwordspacing

\bibitem{Haenggi2016}
M.~Haenggi, ``{T}he meta distribution of the {SIR} in {P}oisson bipolar and
  cellular networks,'' \emph{IEEE Trans. Wireless Commun.}, vol.~15, no.~4, pp.
  2577--2589, Apr. 2016.

\bibitem{Chae2017}
S.~H. Chae, T.~Q.~S. Quek, and W.~Choi, ``{C}ontent placement for wireless
  cooperative caching helpers: {A} tradeoff between cooperative gain and
  content diversity gain,'' \emph{IEEE Trans. Wireless Commun.}, vol.~16,
  no.~10, pp. 6795--6807, Oct. 2017.

\bibitem{Wen2018a}
W.~Wen, Y.~Cui, F.-C. Zheng, S.~Jin, and Y.~Jiang, ``{R}andom caching based
  cooperative transmission in heterogeneous wireless networks,'' \emph{IEEE
  Trans. Commun.}, vol.~66, no.~7, pp. 2809--2825, Jul. 2018.

\bibitem{Feng2021}
T.~Feng, S.~Shi, C.~Wu, and X.~Gu, ``{L}ocation-aware cross-tier cooperation in
  cache-enabled heterogeneous networks,'' in \emph{2021 Int. Wirel. Commun.
  Mob. Comput.}, Jun. 2021, pp. 1748--1753.

\bibitem{Cui2018a}
Q.~Cui, X.~Yu, Y.~Wang, and M.~Haenggi, ``{T}he {SIR} meta distribution in
  {P}oisson cellular networks with base station cooperation,'' \emph{IEEE
  Trans. Commun.}, vol.~66, no.~3, pp. 1234--1249, Mar. 2018.

\bibitem{Feng2019}
K.~Feng and M.~Haenggi, ``{A} location-dependent base station cooperation
  scheme for cellular networks,'' \emph{IEEE Trans. Commun.}, vol.~67, no.~9,
  pp. 6415--6426, Sep. 2019.

\bibitem{Deng2019}
N.~Deng and M.~Haenggi, ``{SINR} and rate meta distributions for {HCNs} with
  joint spectrum allocation and offloading,'' \emph{IEEE Trans. Commun.},
  vol.~67, no.~5, pp. 3709--3722, May 2019.

\bibitem{Wang2019g}
Y.~Wang, M.~Haenggi, and Z.~Tan, ``{SIR} meta distribution of {\$}{K}{\$} -tier
  downlink heterogeneous cellular networks with cell range expansion,''
  \emph{IEEE Trans. Commun.}, vol.~67, no.~4, pp. 3069--3081, Apr. 2019.

\bibitem{Kalamkar2019}
S.~S. Kalamkar and M.~Haenggi, ``{S}imple approximations of the {SIR} meta
  distribution in general cellular networks,'' \emph{IEEE Trans. Commun.},
  vol.~67, no.~6, pp. 4393--4406, Jun. 2019.

\bibitem{Salehi2017}
M.~Salehi, A.~Mohammadi, and M.~Haenggi, ``{A}nalysis of {D2D} underlaid
  cellular networks: {SIR} meta distribution and mean local delay,'' \emph{IEEE
  Trans. Commun.}, vol.~65, no.~7, pp. 2904--2916, Jul. 2017.

\bibitem{Deng2017}
N.~Deng and M.~Haenggi, ``{A} fine-grained analysis of millimeter-wave
  device-to-device networks,'' \emph{IEEE Trans. Commun.}, vol.~65, no.~11, pp.
  4940--4954, Nov. 2017.

\bibitem{Salehi2019}
M.~Salehi, H.~Tabassum, and E.~Hossain, ``{M}eta distribution of {SIR} in
  large-scale uplink and downlink {NOMA} networks,'' \emph{IEEE Trans.
  Commun.}, vol.~67, no.~4, pp. 3009--3025, Apr. 2019.

\bibitem{Yang2022}
\BIBentryALTinterwordspacing
L.~Yang, F.-C. Zheng, Y.~Zhong, S.~Jin, and A.~G. Burr, ``{O}n the {SIR} meta
  distribution for cache-enabled wireless networks with random discontinuous
  transmission: {A}nalysis and optimization,'' \emph{IEEE Trans. Wireless
  Commun.}, vol. 1276, To be published, 2022. [Online]. Available:
  \url{https://ieeexplore.ieee.org/document/9705213/}
\BIBentrySTDinterwordspacing

\bibitem{Wang2018c}
Y.~Wang, Q.~Cui, M.~Haenggi, and Z.~Tan, ``{O}n the {SIR} meta distribution for
  {P}oisson networks with interference cancellation,'' \emph{IEEE Wirel.
  Commun. Lett.}, vol.~7, no.~1, pp. 26--29, Feb. 2018.

\bibitem{Haenggi2012}
M.~Haenggi, \emph{{S}tochastic Geometry for Wireless Networks}.\hskip 1em plus
  0.5em minus 0.4em\relax Cambridge, U.K.: Cambridge Univ. Press, 2012.

\bibitem{Garg2019}
N.~Garg, M.~Sellathurai, V.~Bhatia, B.~N. Bharath, and T.~Ratnarajah,
  ``{O}nline content popularity prediction and learning in wireless edge
  caching,'' \emph{IEEE Trans. Commun.}, vol.~68, no.~2, pp. 1087--1100, Feb.
  2020.

\bibitem{Jiang2019d}
Y.~Jiang, M.~Ma, M.~Bennis, F.-C. Zheng, and X.~You, ``{U}ser preference
  learning-based edge caching for fog radio access network,'' \emph{IEEE Trans.
  Commun.}, vol.~67, no.~2, pp. 1268--1283, Feb. 2019.

\bibitem{Bastug2015}
E.~Baştug, M.~Bennis, M.~Kountouris, and M.~Debbah, ``{Cache-enabled small
  cell networks: modeling and tradeoffs},'' \emph{EURASIP J. Wirel. Commun.
  Netw.}, vol. 2015, no.~1, pp. 1--11, Dec. 2015.

\bibitem{Tamoor-ul-Hassan2015a}
S.~Tamoor-ul Hassan, M.~Bennis, P.~H.~J. Nardelli, and M.~Latva-Aho,
  ``{Modeling and analysis of content caching in wireless small cell
  networks},'' in \emph{ISWCS}, Aug. 2015, pp. 765--769.

\bibitem{Cui2016}
Y.~Cui, D.~Jiang, and Y.~Wu, ``{A}nalysis and optimization of caching and
  multicasting in large-scale cache-enabled wireless networks,'' \emph{IEEE
  Trans. Wireless Commun.}, vol.~15, no.~7, pp. 5101--5112, Jul. 2016.

\bibitem{Wen2017}
J.~Wen, K.~Huang, S.~Yang, and V.~O.~K. Li, ``{C}ache-enabled heterogeneous
  cellular networks: Optimal tier-level content placement,'' \emph{IEEE Trans.
  Wireless Commun.}, vol.~16, no.~9, pp. 5939--5952, Sep. 2017.

\bibitem{Dhillon2013}
H.~S. Dhillon, M.~Kountouris, and J.~G. Andrews, ``{Downlink MIMO HetNets:
  M}odeling, ordering results and performance analysis,'' \emph{IEEE Trans.
  Wireless Commun.}, vol.~12, no.~10, pp. 5208--5222, Oct. 2013.

\bibitem{Li2015a}
C.~Li, J.~Zhang, M.~Haenggi, and K.~B. Letaief, ``{U}ser-centric intercell
  interference nulling for downlink small cell networks,'' \emph{IEEE Trans.
  Commun.}, vol.~63, no.~4, pp. 1419--1431, Apr. 2015.

\bibitem{Hosseini2018}
K.~Hosseini, C.~Zhu, A.~A. Khan, R.~S. Adve, and W.~Yu, ``{O}ptimizing the
  {MIMO} cellular downlink: {M}ultiplexing, diversity, or interference
  nulling?'' \emph{IEEE Trans. Commun.}, vol.~66, no.~12, pp. 6068--6080, Dec.
  2018.

\bibitem{Cui2016a}
Y.~Cui, Y.~Wu, D.~Jiang, and B.~Clerckx, ``{Us}er-centric interference nulling
  in downlink multi-antenna heterogeneous networks,'' \emph{IEEE Trans.
  Wireless Commun.}, vol.~15, no.~11, pp. 7484--7500, Nov. 2016.

\bibitem{GIL-PELAEZ1951}
J.~Gil-Pelaez, ``{Note on the inversion theorem},'' \emph{Biometrika}, vol.~38,
  no. 3-4, pp. 481--482, Jul. 1951.

\bibitem{DimitriP.Bertsekas2016}
{Dimitri P. Bertsekas}, \emph{{Nonlinear Programming 2nd}}.\hskip 1em plus
  0.5em minus 0.4em\relax Belmont, MA, USA: Athena Scientific, 1999.

\bibitem{Zwillinger2014}
{I. S. Gradshteyn and I. M. Ryzhik}, \emph{{Table of Integrals, Series, and
  Products}}.\hskip 1em plus 0.5em minus 0.4em\relax New York, USA: Elsevier,
  2014.

\bibitem{henrici1993applied}
P.~Henrici, \emph{Applied and Computational Complex Analysis, Volume 1: Power
  Series, Integration, Conformal Mapping, Location of Zeros}.\hskip 1em plus
  0.5em minus 0.4em\relax New York, USA: John Wiley \& Sons, 1988.

\bibitem{Li2014}
C.~Li, J.~Zhang, and K.~B. Letaief, ``{T}hroughput and energy efficiency
  analysis of small cell networks with multi-antenna base stations,''
  \emph{IEEE Trans. Wireless Commun.}, vol.~13, no.~5, pp. 2505--2517, May
  2014.

\end{thebibliography}

%
%
%
%
%

\vfill

\end{document}